\documentclass[acmlarge,authorversion]{acmart}

\AtBeginDocument{%
  \providecommand\BibTeX{{%
    \normalfont B\kern-0.5em{\scshape i\kern-0.25em b}\kern-0.8em\TeX}}}

\setcopyright{acmlicensed}
\acmJournal{TOCHI}
\acmYear{2023} \acmVolume{} \acmNumber{} \acmArticle{} \acmPrice{15.00}
\acmDOI{10.1145/3603622}

\usepackage{booktabs}
\usepackage{graphicx}
\usepackage{amsmath}

\newcommand{\figref}[1]{Fig.~\ref{fig:#1}}
\newcommand{\tabref}[1]{Table~\ref{tab:#1}}

\newcommand{\ie}{\emph{i.e.}}
\newcommand{\eg}{\emph{e.g.}}
\newcommand{\etal}{\emph{et al.}}

\begin{document}
%
\title{CLERA: A Unified Model for Joint Cognitive Load and Eye Region Analysis in the Wild}

\author{Li Ding}
\authornote{Work performed when the authors were at Massachusetts Institute of Technology.}
\authornote{Corresponding author.}
\email{liding@umass.edu}
\affiliation{%
  \institution{University of Massachusetts Amherst}
  \city{Amherst}
  \state{MA}
  \country{USA}
  \postcode{01002}
}

\author{Jack Terwilliger}
\authornotemark[1]
\email{jterwilliger@ucsd.edu}
\affiliation{%
  \institution{University of California San Diego}
  \city{La Jolla}
  \state{CA}
  \country{USA}
  \postcode{92093}
}

\author{Aishni Parab}
\authornotemark[1]
\email{aishni@g.ucla.edu}
\affiliation{%
  \institution{University of California Los Angeles}
  \city{Los Angeles}
  \state{CA}
  \country{USA}
  \postcode{90095}
}

\author{Meng Wang}
\authornotemark[1]
\email{mwang0@umass.edu}
\affiliation{%
  \institution{University of Massachusetts Amherst}
  \city{Amherst}
  \state{MA}
  \country{USA}
  \postcode{01002}
}

\author{Lex Fridman}
\email{fridman@mit.edu}
\author{Bruce Mehler}
\email{bmehler@mit.edu}
\author{Bryan Reimer}
\email{reimer@mit.edu}
\affiliation{%
  \institution{Massachusetts Institute of Technology}
  \city{Cambridge}
  \state{MA}
  \country{USA}
  \postcode{02142}
}

\renewcommand{\shortauthors}{Ding, Li et al.}

\begin{abstract}
  Non-intrusive, real-time analysis of the dynamics of the eye region allows us to monitor humans' visual attention allocation and estimate their mental state during the performance of real-world tasks, which can potentially benefit a wide range of human-computer interaction (HCI) applications. While commercial eye-tracking devices have been frequently employed, the difficulty of customizing these devices places unnecessary constraints on the exploration of more efficient, end-to-end models of eye dynamics. In this work, we propose CLERA, a unified model for Cognitive Load and Eye Region Analysis, which achieves precise keypoint detection and spatiotemporal tracking in a joint-learning framework. Our method demonstrates significant efficiency and outperforms prior work on tasks including cognitive load estimation, eye landmark detection, and blink estimation. We also introduce a large-scale dataset of 30k human faces with joint pupil, eye-openness, and landmark annotation, which aims to support future HCI research on human factors and eye-related analysis.
\end{abstract}



\maketitle

\section{Introduction} \label{sec:introduction}

Understanding the appearance and dynamics of the human eye has proven to be an essential component of various human-centered research activities and applications, \eg, visual attention modeling~\cite{borji2012state,ziv2016gaze}, gaze-based human-computer interaction~\cite{majaranta2014eye,fischer2001user,cowie2001emotion}, virtual reality~\cite{patney2016perceptually,clay2019eye}, physical and psychological health monitoring~\cite{harezlak2018application,oliveira2021computer,leveque2018state}, usability evaluation~\cite{goldberg2003eye,joseph2021modeling}, and emotion recognition~\cite{aracena2015neural,lim2020emotion}. However, in order to assess human cognitive load or perform other visual attention modeling tasks in real-world situations, it is often required that the evaluation approach should not interfere with the natural behavior of interest such that the mental state of the individual being measured is not influenced by the measurement approach itself. Moreover, these assessments should also generalize to different environmental and individual-specific characteristics such as visual appearance, movement, pose, scale, perspective, time of the day, etc. Thus, developing practical, non-contact approaches that are not hindered by environmental and experimental constraints remains a challenging problem in HCI research.

Some advanced approaches~\cite{fridman2017can, fridman2018cognitive} have been proposed to take advantage of modern computer vision and deep learning technologies to estimate human cognitive load from an ``in the wild'' perspective through modeling pupil dynamics, which yields the potential of real-time applications such as driver attention monitoring. These approaches often require efficient and precise detection of the eye region and its landmarks, which is often achieved using out-of-the-box eye-tracking devices. However, these devices can be hard to customize and interact with, putting unnecessary constraints on exploring more efficient, end-to-end modeling of eye dynamics. For example, \cite{fridman2018cognitive} needs either the pupil and eye landmark positions or the tracked eye-region image as the input to the cognitive load estimation model. As a result, the current methods are still incapable of making online predictions due to the latency introduced by the prerequisite of adding an eye tracker.

\begin{figure}
  \centering
  \includegraphics[width=\linewidth]{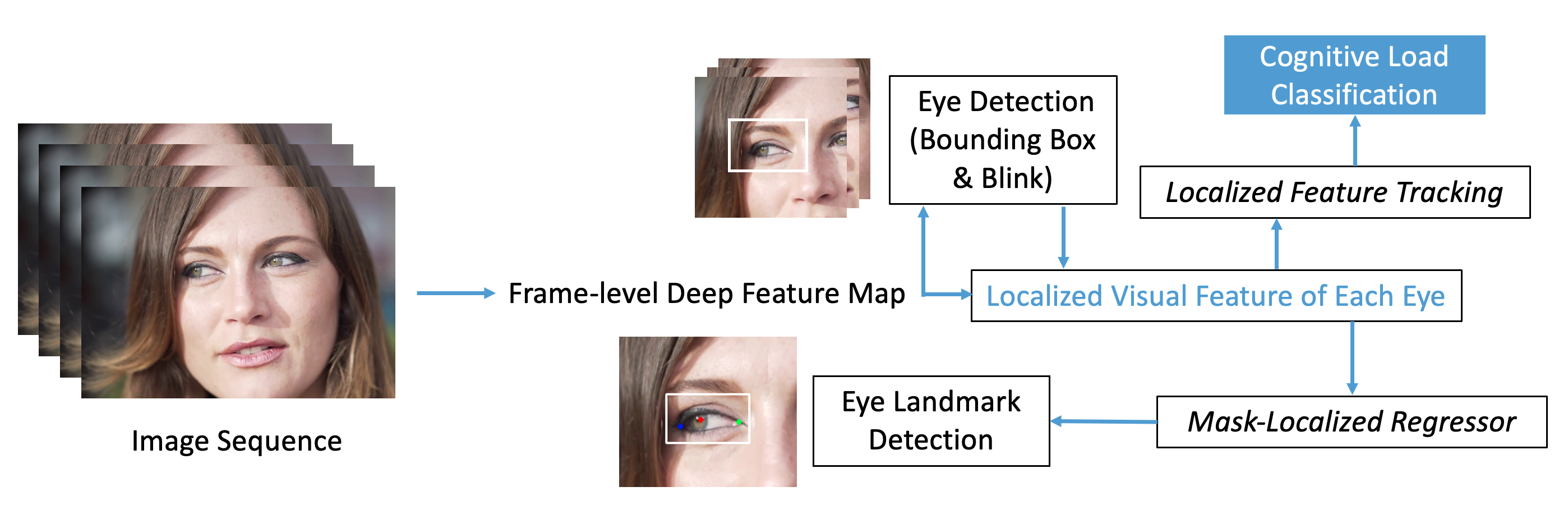}
  \caption{Overview of the proposed model CLERA, for joint cognitive load and eye region analysis. We first perform detection of the eye on the frame-level deep feature map. Next, the architecture extends two heads: \textit{Localized Feature Tracking} for cognitive load estimation over time and \textit{Mask-Localized Regressor} for eye landmark detection.}
  \label{fig:framework}
\end{figure}

In this work, we focus on exploring computer-vision-based joint-learning frameworks for eye-region analysis and downstream eye-dynamics modeling tasks. Our intuition is that since both eye tracking and eye-dynamics modeling tasks can be viewed as learning tasks that take the camera image as input, they could likely share some part of the modeling and be integrated into a joint-learning framework, potentially saving considerable computation for efficiency. To validate this idea, we propose a unified deep learning model, termed \textit{CLERA}, for joint Cognitive Load and Eye Region Analysis, as shown in \figref{framework}.

The proposed model aims to make improvements on prior research from two perspectives. First, we focus on the architectural design of joint-learning deep neural networks. There is very likely to exist a large amount of computation redundancy when the cognitive load estimation model takes image sequences of the eye as input~\cite{fridman2018cognitive}, because the detection of the eye region needs to be predicted by a separate system. As the tasks of eye detection, pupil localization, and cognitive load estimation all rely on extracting visual representations of the human eye region, well-learned representations could potentially be shared across these tasks. Moreover, end-to-end learning of deep neural networks usually requires large-scale data for training. However, obtaining ground truth cognitive load data is usually difficult and costly as it requires specific experimental setups, \eg, n-back tasks~\cite{reimer2012field}. On the other hand, traditional computer vision tasks of eye region analysis, such as eye and landmark detection, are well constructed and easy to annotate at scale. Since cognitive load can be estimated by modeling the physiological reactivity of the eye, successful estimation will need good visual representations of the eye and its components. This fact indicates that eye/landmark detection may serve as side supervision for the cognitive load task to improve the quality of learned representation. Based on these considerations, we propose the \textit{Localized Feature Tracking} technique, which utilizes shared visual features for high-level tasks in the temporal domain, such as cognitive load estimation, within a joint-learning framework. With the detection tasks performed on each frame, we use a temporal tracking algorithm to track each detected eye. For each successfully tracked eye, we perform temporal modeling on the top of localized deep feature maps instead of the raw image. As a result, the whole framework is able to learn general and robust feature representations for precise eye landmarks, and blink detection, and use the same representations for cognitive load estimation, which outperforms existing methods and can efficiently run in real-time to be useful for many real-world applications.

Second, we focus on adapting modern computer vision models to better facilitate real-world applications of eye-region analysis. Existing methods for pupil and blink detection~\cite{lalonde2007real,chen2015real,schillingmann2015yet,lu2014adaptive} heavily rely on the assumption of environmental conditions of the training data, and usually need to work under similar controlled environments. For example, \cite{chen2015real} requires an eye camera to be mounted on eyeglasses, which is not suitable for real-world situations. Recent advancements in common object detection and human pose prediction show exciting performance on large-scale datasets~\cite{lin2014microsoft}.
We leverage this success and frame the pupil and blink detection task as a joint instance and keypoint detection problem. State-of-the-art methods~\cite{he2017mask,chen2018cascaded,papandreou2017towards} tend to use mask-based methods, where the keypoints are predicted using heatmaps on either the whole image or particular regions of interest. The precision of keypoint outputs is thus limited by the resolution of the heatmap. Such approaches are suitable for tasks where the precision of keypoints is not highly demanding as the heatmap resolutions usually suffice, \eg, human pose estimation.
However, when it comes to eye landmarks, mask-based approaches lack the required precision for eye-related tasks, such as capturing the micromovements of the pupil within the eye region. To handle such problems, we propose a method, termed \textit{Mask-Localized Regressor}, that extends mask-based approaches to handle precise eye landmark detection that can provide sub-pixel predictions of coordinates.

In addition, we recognize the need for large-scale datasets to facilitate human factors research using modern data-driven approaches. These approaches often require diverse datasets to capture a variety of natural environments for the task. To meet this need, we propose MIT Pupil Dataset, a large-scale dataset of 30k crowd-sourced web images of human faces. The dataset includes joint annotations for pupil, eye-openness, and landmarks, and has an even distribution of images of closed and open eyes. This dataset aims to serve as an open-source benchmark and to help with the development of modern learning-based algorithms for understanding human eyes in real-world applications. Both the dataset and the algorithm proposed in this work will be released open source to contribute to the community for further research on this topic.

To summarize, the main contributions of this work are:
\begin{enumerate}
  \item \textbf{CLERA}: a unified joint-learning framework for cognitive load and eye region analysis, which consists of two novel techniques:
        \begin{enumerate}
          \item \textit{Localized Feature Tracking} for using shared image features for cognitive load modeling
          \item \textit{Mask-Localized Regressor} for precise eye landmark detection
        \end{enumerate}
  \item \textbf{MIT Pupil Dataset}: a large-scale, open-source\footnote[1]{The dataset is not directly published online due to privacy and sensitivity concerns. For inquiries about using the dataset for research purposes, please contact the corresponding author.} dataset of around 30k images of human faces with joint pupil, eye-openness, and landmark annotation.
\end{enumerate}

\section{Related Work}

\subsection{Cognitive Load Estimation}

The concept of cognitive load~\cite{paas2003cognitive} is often used to refer to the amount of human working memory in use, and has been shown to be an important variable impacting human performance on a variety of tasks, such as machine operations, education, and driving, for which human operators are responsible for the major decision-making and action execution. Early research~\cite{paas2003cognitive,haapalainen2010psycho,sweller2011measuring,mehler2012sensitivity} proposed various physiological measures that are sensitive to changes in cognitive load levels that can be characterized under controlled experimental conditions. There have been numerous studies linking eye movements to variations in cognitive load levels, especially in the driving area. Most studies use different approaches and methods on different datasets. As such, it has been hard to directly compare results. A general finding supported by previous studies is that increased levels of cognitive load often result in a narrowing of visual search space during driving, \ie, gaze concentration. \cite{recarte2003mental} studies the effects of mental workload on visual search and decision-making. \cite{reimer2012field} explores the impact of variations in short-term memory demands on drivers' visual attention and performance. \cite{wang2014sensitivity} compares several methodologies for computing changes in gaze dispersion, showing that horizontal eye movements show the greatest sensitivity to variations in cognitive demand. \cite{zhang2004driver} explores using machine learning methods in driver workload estimation.  \cite{fridman2018cognitive} proposes two novel vision-based methods for cognitive load estimation, and evaluates them on a large-scale dataset collected under real-world driving conditions~\cite{fridman2019advanced}. Our work follows the direction of \cite{fridman2018cognitive} and proposes a more integrated deep learning framework for better efficiency and generalization.  Based on these successful attempts, our work steps further in this direction and integrates the task of cognitive load estimation into the general computer vision task of object detection and video understanding. Such integration allows us to design better deep learning frameworks that manage to improve both computational efficiency and generalization to ``wilder'' real-world circumstances.

\subsection{Pupil Detection and Blink Estimation}
There have been various preliminary works on using image-based computer vision
methods to enhance the safety and experience of driving~\cite{ding2021value,ding2021perceptual,fridman2019arguing,ding2019object}, especially on the driver's facial analysis including detection of human eyes, eye landmarks, gaze, blink, or jointly detect some combination thereof.
Traditional methods~\cite{lalonde2007real,chen2015real,schillingmann2015yet,lu2014adaptive}
either utilize hand-crafted visual features extracted by descriptors such as
SIFT and HOG, or employ image/color models based on the appearance of the human
eye. These methods usually require controlled environments to work, and thus
cannot handle cases in real-world uncontrolled conditions~\cite{fridman2019advanced,ding2020avt,ding2020drivesegm,ding2020drivesegs}, including arbitrary viewpoints, varying face appearances, and illuminance changes. In recent years, deep
learning methods have been used to form better representations in order to
improve the accuracy and robustness of general object
detection~\cite{redmon2016you,ren2015faster,he2017mask} and keypoint
detection~\cite{he2017mask,cao2017realtime,chen2018cascaded,papandreou2017towards}.
Some relevant
papers~\cite{anas2017online,kim2017study,siegfried2019deep,cortacero2019rt}
explore using existing deep learning architectures on the tasks of blink or
gaze estimation. Our work takes a further step to propose a method for precise
keypoint detection and a unified framework designed specifically for joint eye,
pupil, and blink detection. The whole model is optimized for multiple aspects of this task
and enables real-time detection under real-world conditions.

\subsection{Pupil and Blink Datasets}
\tabref{dataset} shows an overview of open-source datasets that have pupil
position and/or eye-openness annotated for real-world images. In general, many
existing
datasets~\cite{tonsen2016labelled,swirski2012robust,fuhl2015excuse,kim2019nvgaze,garbin2019openeds}
for pupil detection and eye tracking involve using head-mounted cameras or
eye-tracking glasses, which are not applicable to many real-world applications
that require practical, non-contact approaches for pupil and blink detection,
and which may also involve localizing eyes of interest in the first place. It is worth noting
\cite{zhang2017mpiigaze} which proposes a larger-scale dataset captured with laptop
webcams for gaze estimation, which also provides pupil and landmark annotations
on a subset. Although this subset has some variability in appearance and
illumination, it still has many constraints such as a limited number of subjects
and camera perspectives. For annotated eye blinks or closed eyes, some existing
datasets~\cite{song2014eyes,pan2007eyeblink,drutarovsky2014eye,fogelton2016eye}
are small in scale in terms of deep network training. \cite{kim2017study}
offers a larger dataset of 5k samples of closed-eye images, but only with
frontal faces. A more recent work~\cite{cortacero2019rt} provides annotations
for a small subset of 480 images with semantic labels of pupil area, and 10k
images of closed eyes. Our dataset aims to address the shortcomings of using
restricted devices, a limited number of subjects, etc., and provides a
large-scale, in-the-wild dataset of around 30k images with joint pupil and eye
landmarks evenly distributed across open and closed eyes.

\section{Methods}

As shown in \figref{framework}, we propose a unified architecture for eye bounding boxes, eye landmarks, blink, and cognitive load estimation with sequential image input. We first extract a frame-level deep feature map using deep convolutional neural networks, and perform bounding box detection and blink estimation (binary classification of open/closed eye). We then locate the positive detection of eyes back onto the feature map and get the localized visual feature representations to perform eye landmark detection. Finally, we track the localized feature through time and use temporal modeling to perform cognitive load estimation.

\subsection{Image Feature Extraction and Eye Detection}

We first use a pre-trained deep convolutional network
(ResNet-50~\cite{he2016deep}) as the image-level feature extractor. ResNet-50 is a deep convolutional neural network architecture that consists of 50 layers, widely used for image recognition tasks. By pre-training on large-scale datasets, it can extract meaningful features from the image to be used for computer vision tasks. On the top
of the feature map (down-scaled by 32 times from the original resolution of the
image), we first perform bounding box detection, which is to predict the bounding box location (coordinates $b_x,  b_y$) and size (width $b_w$, height $b_h$). Instead of following popular
methods~\cite{ren2015faster,redmon2018yolov3} that utilize multi-scale region
proposals, we (similar to \cite{redmon2016you}) predict the bounding box
confidence and its parameters using an extra convolutional layer for better
computation efficiency.

On the feature map, the convolutional layer predicts 5 variables on each cell:
$t_p$, $t_x$, $t_y$, $t_w$, $t_h$. The offset of the cell from the top left
corner of the image is denoted by ($c_x$,$c_y$), and the bounding box prior has
width and height of $p_w$ and $p_h$. The positive predictions (cells that $t_p >
  0$) finally correspond to:
\begin{align}
  b_x & = tanh(t_x)+0.5 + c_x \\
  b_y & = tanh(t_y)+0.5 + c_y \\
  b_w & = p_w\cdot e^{t_w}    \\
  b_h & = p_h\cdot e^{t_h}
\end{align}
Note that our parameterization is different from \cite{redmon2018yolov3} because
in practice, we find the sigmoid used in \cite{redmon2018yolov3} leads to slower
convergence and larger variation in prediction. It also does not allow
predictions to be slightly out of the corresponding cell, which causes accuracy
to decrease when the bounding box happens to locate in the middle of two cells.
When a side cell has higher confidence than the center cell, the non-max
suppression will select the prediction of side cell but it can never predict the
accurate location of the box. Our equation solves this problem by letting each
cell to predict the center of the box at most to the center of the neighboring
cells (the term ($tanh(x)+0.5$) ranges from $-0.5$ to $1.5$). We also perform a
binary classification for each detected eye to obtain its state (open/closed)
for blink detection.

\subsection{Localized Feature Tracking for Cognitive Load Estimation}

With the detection tasks performed on each frame, we use a temporal tracking
method that sets a threshold $\theta$ for the temporal displacement of each of
the detected eye, and track it through time. Namely, given that a detected eye
$(b_{x_t},b_{y_t},b_{w_t},b_{h_t})$ in frame $t$ and another detected eye
$(b_{x_{t+1}},b_{y_{t+1}},b_{w_{t+1}},b_{h_{t+1}})$ in frame $t+1$, we calculate
the Intersection over Unio (IoU) of them,
\begin{equation}
  IoU=\frac{Area\ of\ Overlap}{Area\ of\ Union}= \frac{TP}{TP+FP+FN}
\end{equation}
Where TP is the number of true positives, FP is the number of false positives, and FN is the number of false negatives. If IoU is greater than $\theta$, they are treated as the same eye and tracking is thus established.

For successfully tracked eyes, we further extract the image feature
representation from each frame, and perform temporal modeling on the top of
them. We get the localized feature by locating the feature vector on the feature
map by using the location of the eye. For example, for feature map $l$ that is
16x downsampled from the original resolution, the feature vector of eye
$(b_{x_t},b_{y_t},b_{w_t},b_{h_t})$ will be $l[b_{x_t}/16,b_{y_t}/16]$. The
localized feature vectors of the eye tracked through all the frames are then used as the input for temporal modeling, which outputs estimation of tracked properties such as cognitive load.

For temporal modeling, we use a VGG-like~\cite{simonyan2014very} architecture with blocks of 1D Convolution, BatchNorm~\cite{ioffe2015batch}, and ReLU. Each block consists of 3 Conv-BN-ReLU with the last one having a stride of 2 to perform down-sampling.
The dimension of features in each block is set to [32,64,128,256], and finally
we use global average pooling and a fully-connected layer for classification.
One thing to notice is that the temporal modeling allows gradients to be
back-propagated back to the feature extractor model, which can further fine-tune
the whole model for better performance.

\subsection{Mask-Localized Regressor for Precise Keypoint Detection} \label{sec:mlr}

\begin{figure}
  \centering
  \includegraphics[width=.85\linewidth]{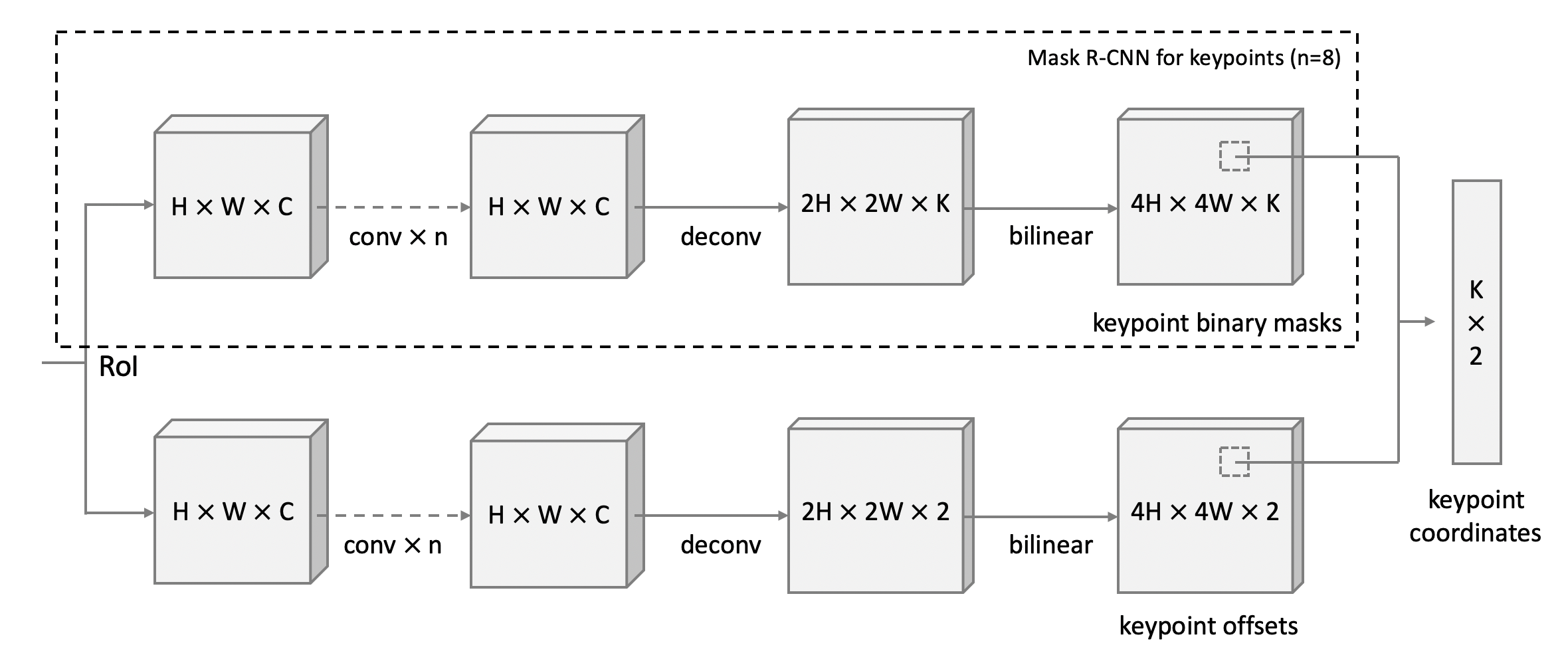}
  \caption{Architecture of the Mask-Localized Regressor Head: We add another
    branch for keypoint offset regression following a similar design as the mask
    branch in \cite{he2017mask}. This branch predicts keypoint-agnostic offsets,
    which are then added to mask-predicted indices to get precise keypoint
    coordinates.}
  \label{fig:head}
\end{figure}

The \textit{Mask-Localized Regressor} is designed to address the limitation of
keypoint coordinate precision in mask-based keypoint detection
approaches~\cite{he2017mask}.
Given a heatmap $f$ of mask coordinates of size $h' \times w'$ corresponding to
the original image or image crop of size $h\times w$, $f(p_i)=1$ if a keypoint
is located at position $p_i$ where $i \in \{1,...,N\}$, $N=h' \cdot w'$. $p_i$
is usually given as a tuple of $(c_x', c_y')$ where $c_x',c_y'$ are rounded
integer coordinates calculated as
\begin{equation}
  (c_x',c_y')^{T} = round\Big((c_x,c_y)^{T}\cdot\frac{(w',h')^{T}}{(w,h)^{T}}\Big),
\end{equation}
where the real coordinates are given as $c_x, c_y$. We then calculate the
coordinate offsets $t_x, t_y$ as
\begin{equation}
  (t_x, t_y)^{T} = \frac{(c_x,c_y)^{T}}{(w,h)^{T}} - \frac{(c_x' + \alpha,c_y'+ \alpha)^{T}}{(w', h')^{T}},
\end{equation}
where $\alpha$ is a fixed offset to adjust index rounding to the grid center,
\eg, $\alpha=0.5$ for zero-based indexing and $\alpha=-0.5$ for one-based.

The Mask-Localized Regressor $g$ models the coordinate offsets $t_x, t_y$
as the lost information during the rounding process, such that $g(p_i)=(t_x,
  t_y)^{T}$ if $f(p_i)=1$. The model predicts on every coordinate $i \in
  \{1,...,N\}$, but only calculates loss if $f(p_i)=1$. So given a loss function
$C(target,prediction)$, the loss is calculated as
\begin{equation}
  loss = \sum_{i=1}^{N} C\Big(g(p_i), (t_x,t_y)^{T}\Big) \cdot f(p_i).
\end{equation}

This method can be integrated into existing frameworks for keypoint detection.
We design an architecture that extends the Mask R-CNN keypoint head to a
Mask-Localized Regressor head, as shown in \figref{head}. We keep the mask
branch as-is, and add a regressor branch with similar architecture for keypoint
offsets. Finally, the two branches are joined together to get precise
keypoint predictions.

Specifically, we perform RoI-Align using the predicted bounding box (ground
truth bounding box during training) on the $8\times$ scale. Both the mask branch and
offset regressor branch have $n=4$, $H=8$, $W=16$, and $C=256$.

\section{MIT Pupil Dataset}

Our goal is to create a dataset suitable to train and evaluate a general-purpose eye detector with the capability to also predict the corresponding attributes of the eye, including pupil, landmark position, and openness. Such a dataset is subject to a few design decisions in order to ensure it covers a sufficiently general domain of scenarios and environmental variations, and can be efficiently and accurately annotated at large-scale. We design a pipeline where we first obtain large-scale images of human faces from different sources to ensure its variability. Then we efficiently annotate the eye and landmarks by splitting the whole annotation process into subtasks: (1) determine if there is a visible right eye present, (2) determine whether the right eye is open or closed, (3) draw a bounding box around a person's right eye, and (4) annotate the keypoints for the right eye. Examples of the dataset are visualized in Figure~\ref{fig:train}.

\subsection{Dataset Structure}
Our dataset is comprised of $28,039$ images, each with the following attributes:
\begin{enumerate}
  \item \textbf{state}: a binary variable, $state \in \{open, closed\}$, which
        marks an eye as either open or closed.
  \item \textbf{bounding\_box}: a quadruple $(x_{bbox1}, y_{bbox1}, x_{bbox2},
          y_{bbox2})$ denoting the upper left and lower right corners of a bounding
        box encompassing an eye.
  \item \textbf{lateral\_canthus}: a tuple $(x_{lateral}, y_{lateral})$ denoting
        the location of the lateral canthus (outside corner) of the eye.
  \item \textbf{medial\_canthus}: a tuple $(x_{medial}, y_{medial})$ denoting
        the location of the medial canthus (inner corner) of the eye.
  \item \textbf{pupil}: a tuple $(x_{pupil}, y_{pupil})$ denoting the location
        of the center of the pupil.
\end{enumerate}

This dataset only includes annotations for the right eyes. By assuming that
differences between the features of the left and right sides of a face vanish at the
population level, the horizontally flipped image of the dataset includes only
left eye annotations. This assumption allowed us to halve the effort needed to
generate the dataset. We describe how we design suitable mechanisms for working with
this dataset for both training and evaluation in Sec.~\ref{sec:singleeye}.

\begin{table*}
  \caption{Overview of open source datasets for annotated pupil and eye
    landmarks position as well as blink/eye-openness. (*: the actual number released for open-source)}
  \label{tab:dataset}
  \centering
  \setlength{\tabcolsep}{.3em}
  \begin{tabular}{lrcrr}
    \toprule
                                                 &                     &                          & \multicolumn{2}{c}{\# of images annotated}                      \\
    \cmidrule(r){4-5}
                                                 & subjects            & camera view              & pupil \& landmarks                         & blink / closed eye \\
    \midrule
    Swirski \etal~\cite{swirski2012robust}       & 2                   & head-mounted             & 600                                        & -                  \\
    Fuhl \etal~\cite{fuhl2015excuse}             & 17                  & head-mounted             & 38,401                                     & -                  \\
    LPW~\cite{tonsen2016labelled}                & 22                  & head-mounted             & 130,856                                    & -                  \\
    MPIIGaze~\cite{zhang2017mpiigaze,zhang2017s} & 15                  & frontal (laptop)         & 37,667 (10,848*)                           & -                  \\
    NVGaze~\cite{kim2019nvgaze}                  & 3                   & head-mounted             & 7,128                                      & -                  \\
    OpenEDS~\cite{garbin2019openeds}             & 152                 & head-mounted             & 12,759                                     & -                  \\
    ZJU~\cite{pan2007eyeblink}                   & 20                  & frontal \& upward        & -                                          & 255 / 1,016        \\
    Kim \etal~\cite{kim2017study}                & -                   & frontal                  & -                                          & - / 4,891          \\
    CEW~\cite{song2014eyes}                      & 2,423               & wild (internet)          & -                                          & - / 1,192          \\
    Eyeblink8~\cite{drutarovsky2014eye}          & 4                   & frontal                  & -                                          & 353 / -            \\
    Res. Night~\cite{fogelton2016eye}            & 107                 & frontal (screen)         & -                                          & 1,849 / -
    \\
    RT-BENE~\cite{fischer2018rt,cortacero2019rt} & 15                  & free-viewing             & 480                                        & -
    / 10,444                                                                                                                                                        \\
    \midrule
    \textbf{Ours}                                & \textbf{$>$10,000}  & \textbf{wild (internet)} &
    \textbf{24,391}                              & \textbf{- / 13,764}                                                                                              \\
    \bottomrule
  \end{tabular}
\end{table*}

\subsection{Image Collection}

We assemble a collection of approximately 30k images that consists of a wide
variety of faces with equal instances of closed and open eyes. We used several
existing datasets to compile a preliminary set of around 7k images, including
Labeled Faces in the Wild~\cite{huang2008labeled}, CAS-PEAL~\cite{gao2007cas},
Caltech Faces 1999~\cite{faces1999database} and Closed Eyes in the
Wild~\cite{song2014eyes}. To collect more closed-eye images, we used search
engines to locate open-source licensed real-world faces varying in head pose, gaze,
race, gender and lighting conditions. These provided 17k images. In order to
capture the intrinsic pupil movement of the human eye, we also gathered a set of
high-resolution YouTube videos, and captured 6k images from those videos.

\subsection{Annotation Process}

For all annotation tasks, we employed professional in-house annotators and
developed web-based custom annotation tools in order to produce high-quality and
efficient annotations. The annotators are trained with a warm-up task of 100 or more images and required to pass the manual check by researchers to start annotation.

\subsubsection{Bounding Box Annotator}

\begin{figure}
  \centering
  \includegraphics[width=0.43\textwidth]{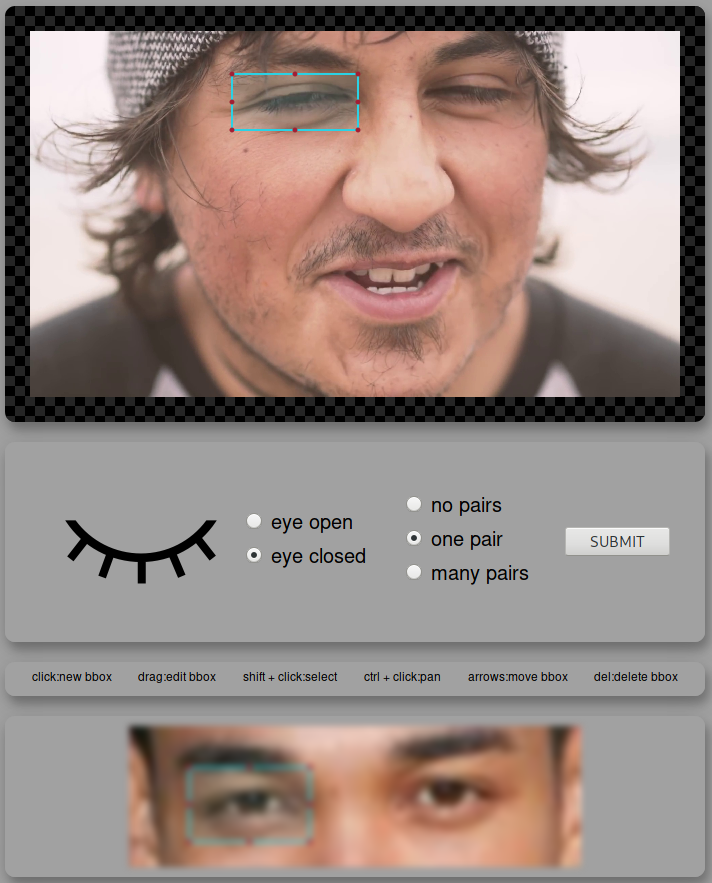}
  \includegraphics[width=0.48\textwidth]{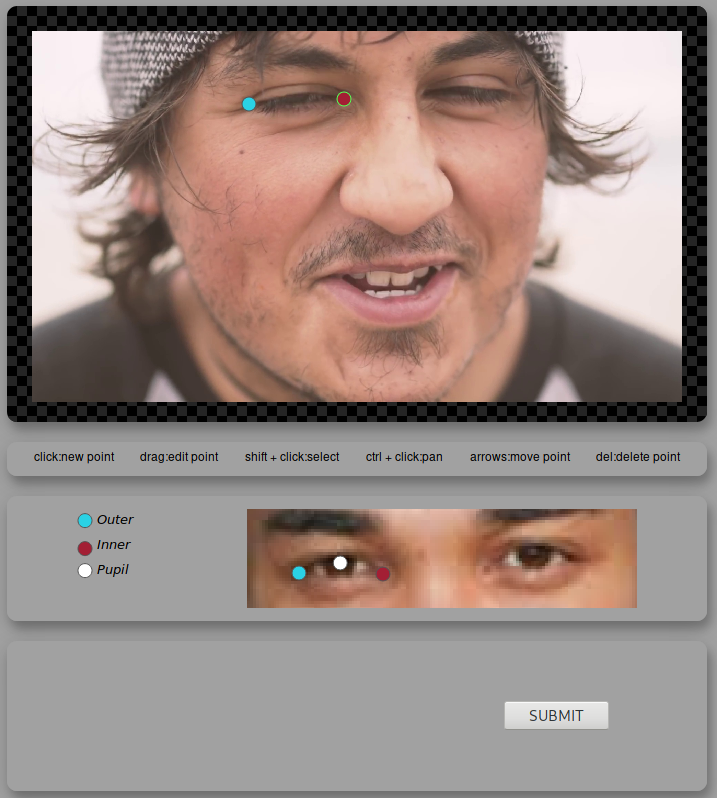}
  \caption{The Bounding Box and Keypoint Annotation Interfaces}
  \label{fig:interface}
\end{figure}

The bounding box interface, shown in \figref{interface} on the left, shows the
user one candidate image at a time. First, the annotators are asked to tell how
many pairs of eyes exist in the image. If there is exactly one pair of eyes in
the image (including occluded), the annotators go on to draw a bounding box
around the right eye and select whether the eye is open or closed. The
interface allows annotators to zoom in, drag the box, or move it by small steps
using the arrow keys in order to make annotation easier and more accurate.

\subsubsection{Keypoint Annotator}

As shown in \figref{interface} on the right, the keypoint interface shows the
annotator an image that has been annotated with a bounding box of the right eye,
and asks workers to click on the locations of eye landmarks of interest. Annotators first annotate the lateral canthus, then the medial canthus, and lastly the pupil. The annotated locations
of these landmarks are shown with corresponding colors indicated in \figref{interface}.

We found that by enforcing the order of landmarks to be annotated as prescribed
by the interface helps to improve efficiency and eliminates errors associated with missing
the order of landmarks: an annotator merely has to click their mouse three times to
annotate an image as opposed to manually selecting a landmark type before each
click. Annotators are also able to adjust the location of annotated landmarks by
either dragging or using the arrow keys.

We also have annotators provide a redundant label for each eye state, which helps
us verify the quality of annotation. This is done by asking annotators to annotate
the pupil only if they (1) believe the eye is open and (2) the center of the
pupil is visible or inferrable.

\subsubsection{Two-pass Annotation}

The annotation occurred in two passes, in which each image was annotated by two
different annotators. This allowed us to measure annotation integrity and
accuracy by comparing the agreement and disagreement between the two
annotations. We then perform multiple experiments to determine suitable filters
for each of the tasks to programmatically remove either false or ambiguous
annotations from the dataset to improve the overall data quality, as described
in the next section.

\begin{table}
  \caption{Two-pass keypoint annotation statistics.}
  \label{tab:kptanno}
  \centering
  \setlength{\tabcolsep}{.8em}
  \begin{tabular}{lcccc}
    \toprule
                    & \multicolumn{4}{c}{Euclidean distance (normalized by box width)}
    \\
    \cmidrule(r){2-5}
    Keypoint        & $<0.05$                                                          & $[0.05, 0.1)$ & $[0.1, 0.2)$ & $\geq 0.2$ \\
    \midrule
    Pupil           & $82.8\%$                                                         & $9.1\%$       & $1.0\%$      & $7.1\%$    \\
    Lateral Canthus & $46.7\%$                                                         & $30.2\%$      & $16.9\%$     & $6.2\%$    \\
    Medial Canthus  & $52.5\%$                                                         & $23.2\%$      & $10.0\%$     & $5.3\%$    \\
    \bottomrule
  \end{tabular}
\end{table}

\subsection{Data Validation}

With the two-pass annotation, we measure the consistency in the annotations, \ie, the degree to which annotations performed on the same image by two different
workers agreed for bounding box, eye state, and keypoints.

\subsubsection{Bounding Box} We first exclude all the images with less than two
bounding box annotations. Then we use a common similarity metric, Intersection
over Union (IoU), to compare bounding box annotations from both passes. The
average IoU is $88.0\%$. We also observed that the bounding boxes have larger
variations horizontally than vertically. The ambiguity in height of the eye
caused the majority of inconsistencies in annotations.

\subsubsection{Eye State} After bounding box filtering, we compare the eye
state labels from two workers. $93.5\%$ of the annotations agree with each
other, and, for the rest, we labeled those as ambiguous cases and excluded them from the
current dataset.

\subsubsection{Keypoints} We compare keypoint annotations by measuring the
Euclidean distance between the coordinates from two-pass annotations for each of
the three keypoints: lateral canthus, medial canthus, and pupil. The keypoint
distance is measured in pixels and then scaled to the corresponding bounding box
width, which is more consistent than box height or area. The statistics are
shown in Table~\ref{tab:kptanno}. We observe that pupil annotations are more
consistent than those for lateral and medial canthus. Note that for distance $\geq 0.2$,
it also includes situations when one of the workers did not annotate the
keypoint, which happens more often in pupil annotation cases where one of the workers
rates the pupil as not visible.

\subsubsection{Final Dataset}
Finally, we also add filters to remove images that are out of the
area of interest in this work. For the final dataset, we applied the following filters
to automatically clean up the dataset:
\begin{enumerate}
  \item \textbf{bounding box}: Removed all images where bounding box IoU $<
          0.3$, which we consider as cases where two annotators disagree with each other, \eg, they annotate different eyes in the image. We then take mean box coordinate values.
  \item \textbf{state}: removed all images where eye state annotations disagree.
  \item \textbf{keypoints}: Removed all keypoint annotations where any of the
        three normalized keypoint distances $\geq 0.2$, which we consider as two annotators disagree. The box and state annotations are still kept if keypoint annotation is removed. We take mean keypoint coordinate values for the rest.
  \item \textbf{out-of-interest}: Removed all cases where either the images have
        low resolution (width of the eye bounding box less than 30 pixels) or rotated
        over 45 degrees (inferred from keypoint positions).
\end{enumerate}

\section{Experiments}

In this section, we describe the emprical studies that consist of
multiple tasks including cognitive load estimation, eye landmark detection, and
blink detection. Since no prior work is evaluated on all these tasks, we
evaluate our model separately on each task for better comparison.

\subsection{Cognitive Load Estimation}

Varied findings have been reported in the literature regarding the responsiveness of gaze concentration measures to changes in cognitive demand. We hereby describe the experiments and results regarding the validation of the proposed method on the cognitive load estimation task.

\subsubsection{Dataset}

In this experiment, we use an unpublished dataset (obtained and extended from
\cite{reimer2012field}) of 212 30-second video clips of driver faces, each under
one of two different cognitive load levels (104 low and 108 high), across 81
different subjects. The subjects needed to meet the criteria of being proficient and regular drivers, which was defined as having a valid driver's license for at least three years and driving a minimum of three times per week. The data were collected with a Volvo XC90 vehicle, which was equipped with synchronized data collection capabilities from a range of built-in sensors including vehicle's controller area network (CAN), cameras for recording driver behavior and the surrounding environment, and audio captured from within the vehicle cabin. The study utilized a delayed digit-recall task, known as the n-back task, with three distinct levels of difficulty to impose varying degrees of secondary cognitive workload on the drivers. In this work, we use the data from low and high levels to form a binary classification task.

Previous research on cognitive load estimation has primarily utilized synthetic or controlled environments, such as driving simulator~\cite{liang2007real,pedrotti2014automatic}, tele-surgical robotic simulation~\cite{wu2020eye}, and simulation games~\cite{appel2021cross}, limiting their applicability to real-world situations. However, our work is focused on addressing this limitation by investigating the estimation of cognitive load in real-world settings. While recent work~\cite{fridman2018cognitive} has explored using real-world testing cases, we take one step further and use a significantly more challenging dataset with varying lighting conditions and camera placements. By conducting our research in naturalistic environments, we aim to capture the complexity and variability of real-world cognitive load scenarios, which could provide valuable insights for enhancing the development of cognitive load estimation models that can be applied in practical settings.

\subsubsection{Comparison Methods}

We use the same experimental settings as in \citet{fridman2018cognitive} that average the results over 10 random training/testing splits (80\% for training and 20\% for testing) across subjects. For comparisons, we first implement the SVM approach in \citet{liang2007real} to serve as the baseline method. We also implement the HMM approach in \cite{fridman2018cognitive}, which is one of the state-of-the-art approach using pupil position to estimate driver's cognitive load.

\begin{table}
  \centering
  \caption{Results for the cognitive load classification task. Our proposed
    method outperforms previous work and shows the significance of using the
    localized feature tracking.}
  \begin{tabular}{l|c}
    \toprule
    Method                                                      & Classification Accuracy \\
    \midrule
    Eye Feature + SVM~\cite{liang2007real}                      & 59.43\%                 \\
    Horizontal Pupil Position + HMM~\cite{fridman2018cognitive} & 61.97\%                 \\
    CLERA (w/ Horizontal Pupil Position)                        & 63.43\%                 \\
    CLERA (no fine-tune)                                        & 64.81\%                 \\
    \textbf{CLERA}                                              & \textbf{66.58\%}        \\
    \bottomrule
  \end{tabular}
  \label{tab:cog}
\end{table}

\subsubsection{Results}

The results are shown in Tab.~\ref{tab:cog}. First, we implement the HMM model
with horizontal pupil position from \cite{fridman2018cognitive} as a baseline.
To validate the effectiveness of each component in CLERA, we implement two
variants of CLERA: CLERA (w/ Horizontal Pupil Position) is using horizontal
pupil position as input to the temporal modeling, and CLERA (no fine-tuning) is
using the proposed localized feature tracking without fine-tuning the feature
extractor for temporal modeling.

We can first observe that CLERA (w/ Horizontal Pupil Position) outperforms the
HMM when using the same horizontal pupil position as the input. This result
aligns well with the observation in prior work~\cite{fridman2018cognitive} where
the 3D-CNN model outperforms the HMM. Secondly, CLERA using the proposed
localized feature tracking outperforms the one using horizontal pupil position.
This indicates that there is information loss when abstracting eye movement to the change in normalized pupil position, and the localized feature can be
used as a better feature with minimal extra computation cost. Thirdly, since
CLERA allows end-to-end gradient learning, the full CLERA model gets a large
performance gain, which suggests that the cognitive load task needs some
specific visual representation that can not be learned from other vision tasks
such as eye landmark detection.

In general, all the CLERA variants are able to outperform prior work, and the
full model has a significant improvement. It is worth noting that while the
absolute accuracy values obtained for differentiating the two cognitive load
states was moderate, the test dataset consisted of data collected in a moving
environment (a vehicle), with individuals having variable positioning relative to the camera and under variable lighting conditions - a very challenging real-world dataset as oppressed to data collected under controlled laboratory conditions~\cite{liang2007real,pedrotti2014automatic,wu2020eye,appel2021cross}. For example, \cite{liang2007real} claims to have over 80\% accuracy in detecting driver cognitive distraction, but only have below 60\% accuracy in our evaluation, which indicates that there exists a considerable difference between simulated and real-world environments, and more future efforts are required to address this issue.

Nevertheless, the primary interest for this work is to show the increase in performance across the proposed methods. We evaluate some of the broader capabilities of our methods in the next sub-section.

\subsection{Eye State and Landmark Detection}
\label{sec:landmark}

\subsubsection{Metrics}

We adopt the Average Precision (AP) metrics for the eye localization task, which is an evaluation metric commonly used in machine learning to measure the accuracy of object detection or segmentation models. It is calculated by computing the area under the Precision-Recall curve (AP-PR) for a given set of predictions and ground truth labels. The formula for calculating AP is as follows:
\begin{equation}
  AP = \frac{\sum_n (R_n - R_{n-1}) \cdot P_n}{R_{tot}}
\end{equation}
where $P_n$ and $R_n$ are the precision and recall at the $n$th threshold, $R_{tot}$ is the total number of positive examples, and $R_{n-1}$ is the recall at the previous threshold.

As we observe evident variation in bounding box annotations,
we mainly focus on keypoint metrics for evaluation, but use box metrics for
ablation experiments on the detection backbone. For blink/eye-openness
prediction, we simply calculate the accuracy, since the MIT Pupil Dataset is
well-balanced for both cases. In terms of keypoint detection, we follow a similar
principle as the existing OKS metric~\cite{lin2014microsoft}, and propose a
specific metric for eye landmarks. We first measure the Euclidean distance
between ground truth and predicted eye landmarks, normalized by the width of
corresponding bounding boxes, which is the same metric as we used in
\tabref{kptanno}. To calculate the AP on this distance for a range of levels, we
choose to use more standard and interpretable level definitions as simply
$.01:.01:.1$. To address the problem that lateral and medial canthus have a larger variance than the pupil (approximately $2\times$), we add a factor of $0.5$ to the two and
calculate the weighted mean for all three landmarks (two if closed eye) for AP
calculation, meaning all the keypoints are jointly evaluated together for each
detection.

In order to work with single-eye annotation, we add the following rule to the AP
calculation process: ignore the first detection if it has no overlap with ground
truth (for bounding box), or the weighted distance is greater than 0.5 (for
keypoints). This is because we do not want to count for potentially correct
detection for the other eye. For the training and testing split, we perform a random
8:2 split and use the same dataset split for all the experiments.

\subsubsection{Baseline Methods} \label{sec:baseline}

Since no prior work has been done on the proposed dataset, we also propose three
methods for the benchmark and ablation study. In order to make fair comparisons, we
use the same detection head as in Sec.~\ref{sec:mlr} and focus on validating
keypoint detection performance. The first method, called
\textit{Baseline-Regressor}, adds another regressor subnet along with two
subnets in the RetinaNet detection head, which has the same subnet architecture
and directly predicts the offsets of each keypoint for each anchor box. The
second method, called \textit{Baseline-Mask}, simply uses the mask branch in
the mask-localized regressor head, which can be viewed as an implementation of Mask R-CNN~\cite{he2017mask} with our specific backbone. This method is only optimized on mask loss. The third method for
comparison, CLERA (Mask-only), uses the proposed Mask-Localized Regressor
jointly trained on mask loss and offset regression loss, but only uses the mask
predictions without adding offset predictions. This is intended to show the
direct performance of offset prediction. With the above three methods, we can
more clearly separate and show the improvement gained from using the proposed
Mask-Localized Regressor.

\subsubsection{Working with Single-Eye Annotation} \label{sec:singleeye}

\begin{figure}
  \includegraphics[width=\linewidth]{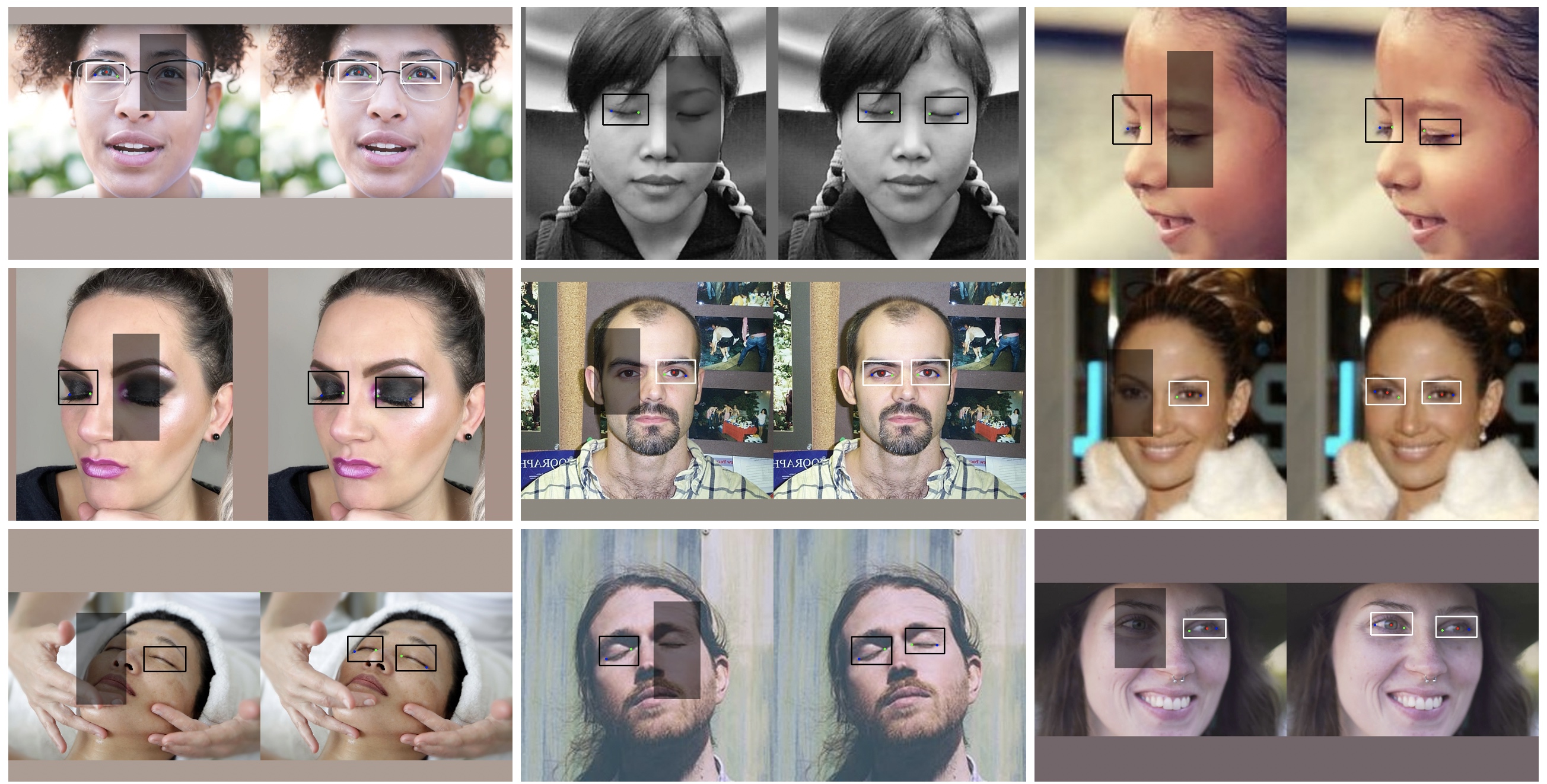}
  \includegraphics[width=\linewidth]{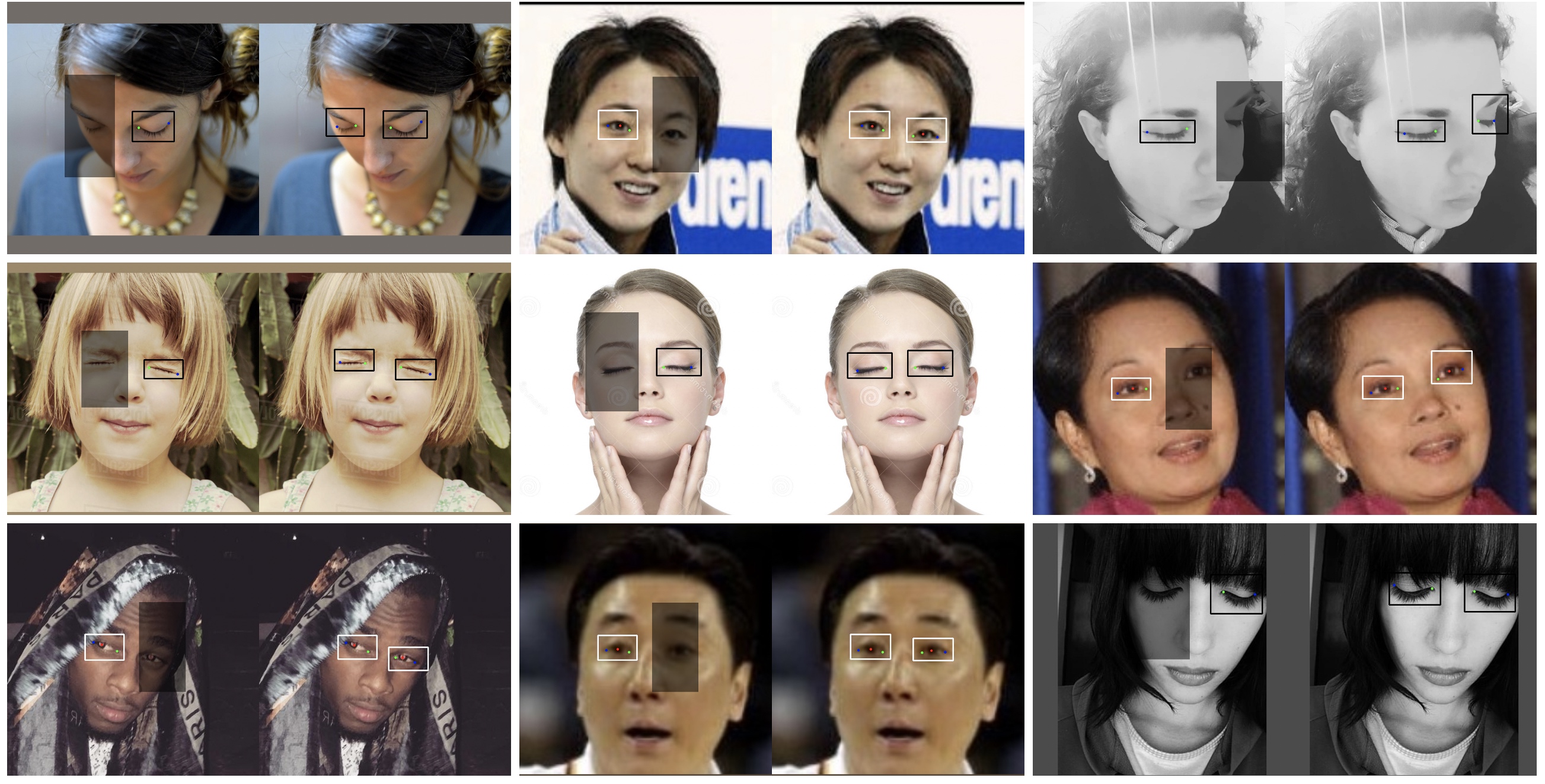}
  \caption{Sample visualization of pairs of ground truth annotations (on the
    left) and predictions of CLERA (on the right) on the training set of
    MIT Pupil Dataset. The color of bounding boxes indicates the eye state (white for
    open and black for closed). The eye landmarks are visualized as colored dots
    (blue for lateral canthus, green for medial canthus, and red for pupil). The
    shadowed area indicates the gradient masking we applied during training with
    single-eye annotation, which is the potential area of the center of the left eye
    (or right eye if horizontally flipped) that we do not annotate.}
  \label{fig:train}
\end{figure}

Since the proposed MIT Pupil Dataset only contains annotation of the right eye, in
order to make the detector also capable of detecting the left eye, we use a
training strategy with horizontal flipping and inferred gradient masking. During
the training process, we first infer the region where potentially the other eye
exists by using the position of the known eye, and generating a mask for that
region. During training, that region is ignored for loss calculation in eye
localization, as we do not have enough information to evaluate or penalize the
detections in that region. As we add the horizontally-flipped version of the
input image and corresponding annotations to generate left-eye samples, the
model finally converges to detect both right eyes and left eyes at the same
time. We visualize the prediction of CLERA on the training set in Figure~\ref{fig:train}, which demonstrates that the model learns to predict both eyes using this training strategy.

\begin{table}
  \caption{Eye state and landmark detection results on MIT Pupil Dataset.}
  \label{tab:expkpt}
  \centering
  \setlength{\tabcolsep}{.5em}
  \begin{tabular}{c|cccccc}
    \toprule
    Methods            & mAP           & AP.1          & AP.04         & AP.02 & State Acc. & FPS           \\
    \midrule
    Baseline-Regressor & 53.7          & 76.2          & 55.7          & 7.5   & 98.6       & \textbf{38.7} \\
    Baseline-Mask      & 69.1          & 87.5          & 77.4          & 28.5  & 98.7       & 38.5          \\
    CLERA (mask-only)  & 70.8          & 90.7          & 79.5          & 28.9  & 98.7       & 38.5          \\
    \textbf{CLERA}     & \textbf{71.1} & \textbf{90.7} & \textbf{79.8} &
    \textbf{30.4}      & \textbf{98.7} & 38.3                                                               \\
    \bottomrule
  \end{tabular}
\end{table}

\subsubsection{Results}

\tabref{expkpt} shows the experimental results for eye state and landmark
detection. Since the models are jointly trained for multiple tasks, we increase
the number of training iterations to 80k and the batch size to 32 for better
convergence. We compare the proposed method to the comparison methods as
described in Sec.~\ref{sec:baseline}. The overall results show that while the
eye state accuracy stays similar, our method significantly outperforms the
other methods on the landmark prediction task.

To dive deeper into the results, first, by comparing Baseline-Mask with CLERA
(mask-only), it shows that adding the offset prediction branch helps the joint
model to learn better mask predictions; secondly, by comparing the full CLERA
model with CLERA (mask-only), while the loose metrics (AP.1) stay the same, we
observe consistent improvements on strict metrics (AP.04 and more significantly
AP.02), and also on overall performance (mAP). This result aligns well with our
intuition in proposing the Mask-Localized Regressor for more precise keypoint
predictions, which can be better evaluated with strict metrics.

The proposed model is also efficient and runs in real time. The FPS is
calculated over the whole testing set. We benchmark all of the runtime results
using the same desktop machine with Nvidia 1080Ti GPU, and the inference is carried out
with one image per batch, with max dimension rescaled to 512.

\subsection{Cross-dataset Evaluation}

\subsubsection{Eye Landmark Detection}
While there are no similar open-source datasets for real-world non-contact eye
landmark detection in unconstrained environments (Table~\ref{tab:dataset}), we adopt the MPIIGaze~\cite{zhang2017mpiigaze,zhang2017s} dataset, which
also features eye landmark annotation but with a limited set of subjects and
environments, and create an external testing set for eye landmark detection.
More specifically, we jointly use the face images provided in \cite{zhang2017s}
and the eye landmark annotation provided in \cite{zhang2017mpiigaze}, resulting in a dataset of 3,877 face images with annotated eye landmarks for pupil, lateral and medial canthus for both eyes.

We perform similar experiments as described in Sec.~\ref{sec:landmark}, using the same models trained on the MIT Pupil Dataset training set and evaluate on
this subset of MPIIGaze dataset. Since there is no bounding box annotation
provided, we normalize the errors with the distance between the corners by a
factor of $1.3$ as an alternative to box width. The results are shown in
\tabref{expkptmpii}.

First, we observe similar overall results showing that the proposed model
consistently outperforms the other methods on the landmark prediction task.
However, the improvements on strict metrics (AP.04 and AP.02) are not as
significant compared to the results in Table~\ref{tab:expkpt}. In
addition, while the same models perform markedly better on MPIIGaze under loose
metrics (AP.1) than on the MIT Pupil Dataset, suggesting that MPII is an easier
benchmark because of its constraints, the results for strict metrics are
actually the opposite. After further investigation, we conclude that this
is because the MPIIGaze dataset is of a lower resolution and the landmark
annotations are rounded to integer. As a result, MPIIGaze is not sufficient for
evaluating keypoints at high precision. We suggest future work adopt the
MIT Pupil Dataset for better evaluation of eye landmark detection in terms of both
precision and robustness.

\begin{table}
  \caption{Eye landmark detection results on the MPIIGaze dataset.}
  \label{tab:expkptmpii}
  \centering
  \setlength{\tabcolsep}{.5em}
  \begin{tabular}{c|cccc}
    \toprule
    Methods            & mAP           & AP.1          & AP.04         & AP.02 \\
    \midrule
    Baseline-Regressor & 58.5          & 86.6          & 54.0          & 11.1  \\
    Baseline-Mask      & 64.0          & 94.7          & 56.2          & 11.4  \\
    CLERA (mask-only)  & 65.1          & 98.3          & 57.7          & 11.2  \\
    \textbf{CLERA}     & \textbf{65.5} & \textbf{98.4} & \textbf{58.9} &
    \textbf{11.7}                                                              \\
    \bottomrule
  \end{tabular}
\end{table}

\subsubsection{Blink Detection}
We also evaluate the performance of the proposed method on the RT-BENE
dataset~\cite{cortacero2019rt} as the testing set, which has large-scale blink
annotation but with a limited set of subjects, and compare the results to
existing methods. We directly use the face images provided in
\cite{fischer2018rt} instead of the cropped eye images in
\cite{cortacero2019rt}. Since the images are of lower-resolution at
224$\times$224, we apply rescaling to 448$\times$448.

The blink prediction is obtained as the predicted state of one detected eye with
the highest confidence on each image. We use the whole RT-BENE dataset with 114,490
images. The blink evaluation is only performed on images with at least one
detected eye, which corresponds to $99.8\%$ of all the samples. The results are
shown in \tabref{expblink}.

\begin{table}
  \caption{Blink detection results on the RT-BENE dataset. (The FPS of CLERA is
    calculated for running the full model for joint eye, blink, and landmark
    detection on a single Nvidia 1080Ti GPU.)}
  \label{tab:expblink}
  \centering
  \setlength{\tabcolsep}{.5em}
  \begin{tabular}{c|ccccc}
    \toprule
    Method                                       & Precision      & Recall         & AP             & F1             & FPS            \\
    \midrule
    Google ML-Kit~\cite{cortacero2019rt}         & 0.172          & \textbf{0.946} & 0.439          & 0.290          & –              \\
    Anas \etal~\cite{anas2017online}             & 0.533          & 0.537          & 0.486          & 0.529          & \textbf{408.3} \\
    RT-BENE - MobileNetV2~\cite{cortacero2019rt} & 0.579          & 0.604          & 0.642          & 0.588          & 42.2           \\
    RT-BENE - ResNet~\cite{cortacero2019rt}      & \textbf{0.595} & 0.610          & 0.649          & 0.598          & 41.8           \\
    \midrule
    \textbf{CLERA - ResNet}                      & 0.571          & 0.750          & \textbf{0.653} & \textbf{0.648} & 42.6           \\
    \bottomrule
  \end{tabular}
\end{table}

Comparing the RT-BENE models that require cropping of the eye region beforehand and
are computationally heavy for only the blink classification task, our model
(with ResNet-101 backbone) not only shows competitive performance on blink
detection, but more importantly, it is a single model that handles joint eye,
blink, and landmark detection in real-time with input at a 2X higher
resolution. The results suggest that the proposed model successfully utilizes the
shared deep features for multiple tasks. It also shows the generalization of
models trained on MIT Pupil Dataset that can be applied on other datasets directly with
promising performance.

\subsubsection{Qualitative Results}

\begin{figure}
  \centering
  \includegraphics[width=\linewidth]{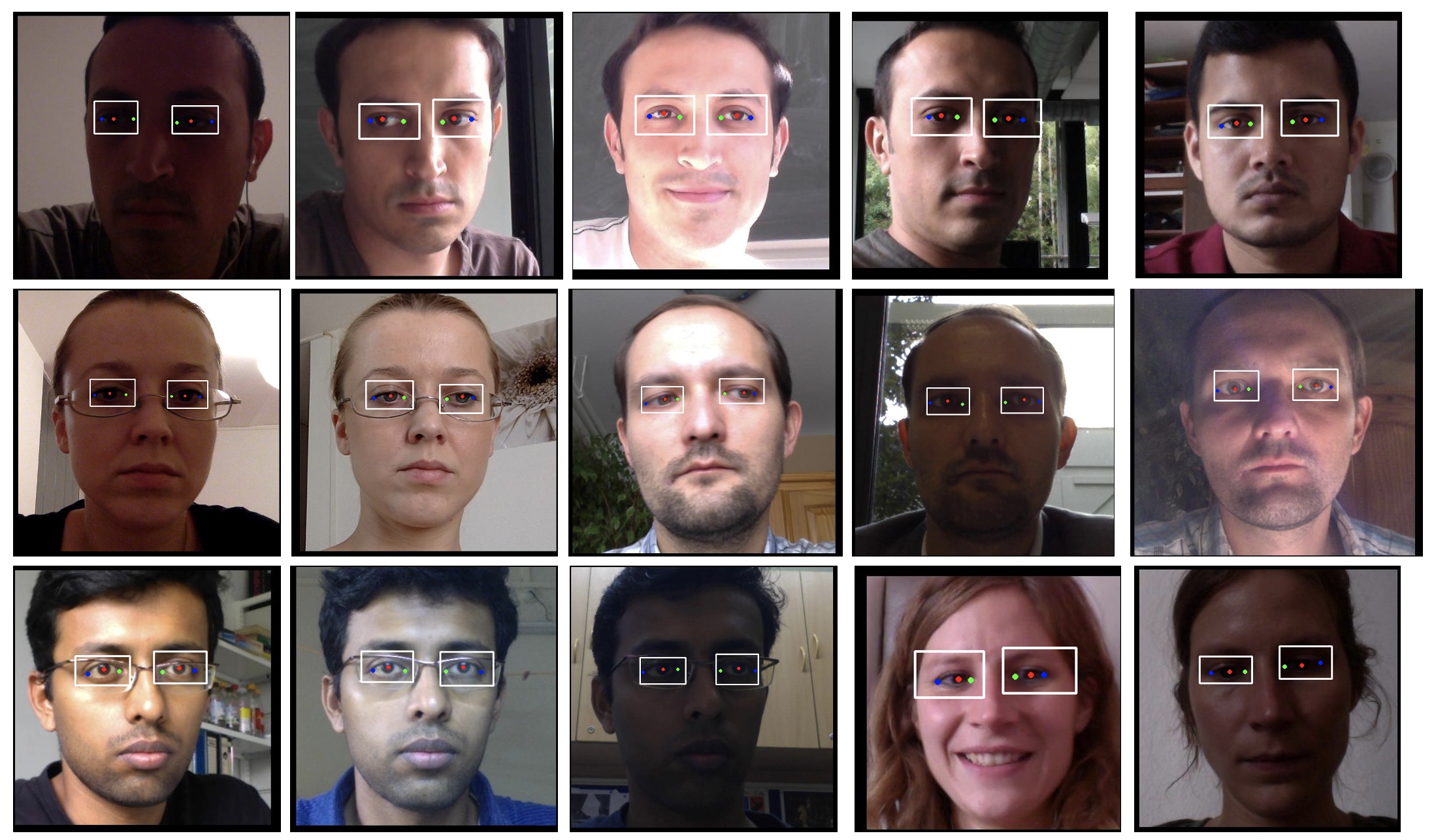}
  \caption{Sample visualization of predictions on MPIIGaze dataset.}
  \label{fig:mpii}
\end{figure}

\begin{figure}
  \centering
  \includegraphics[width=\linewidth]{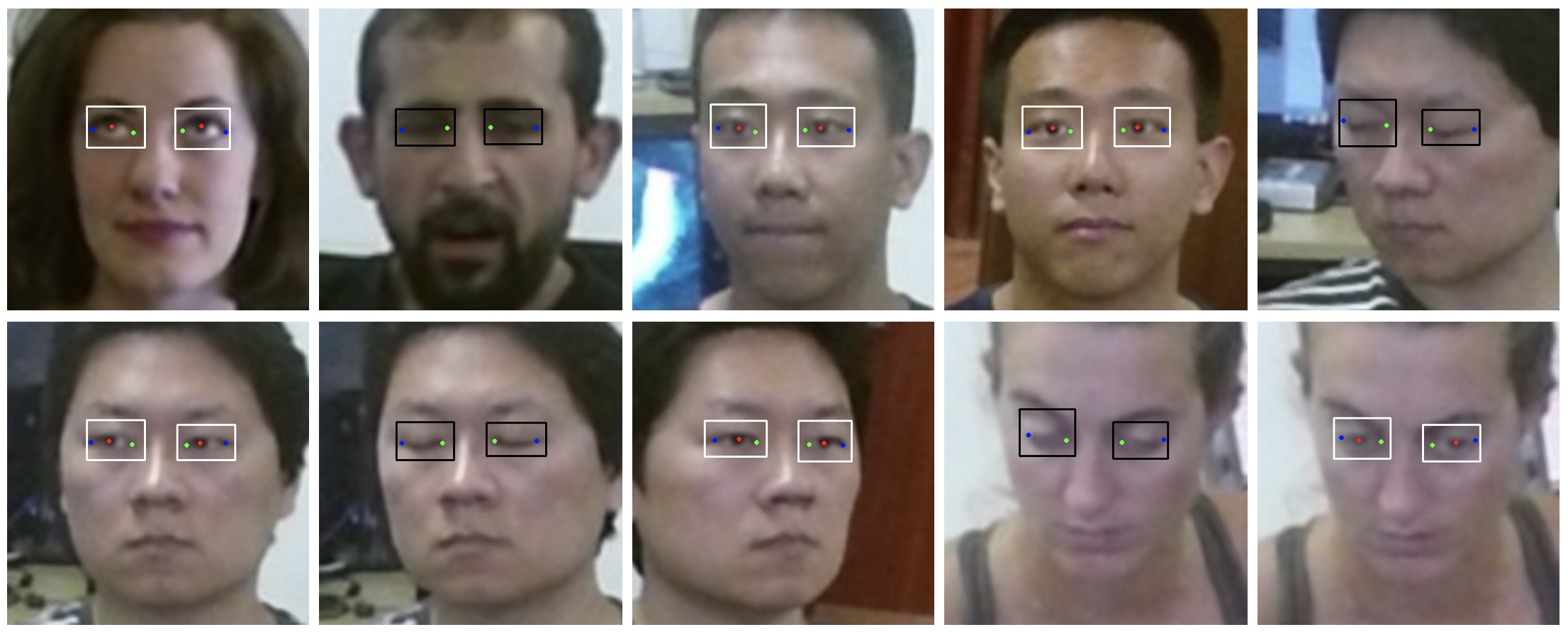}
  \caption{Sample visualization of predictions on RT-BENE dataset.}
  \label{fig:rt}
\end{figure}

\begin{figure}
  \centering
  \includegraphics[width=\linewidth]{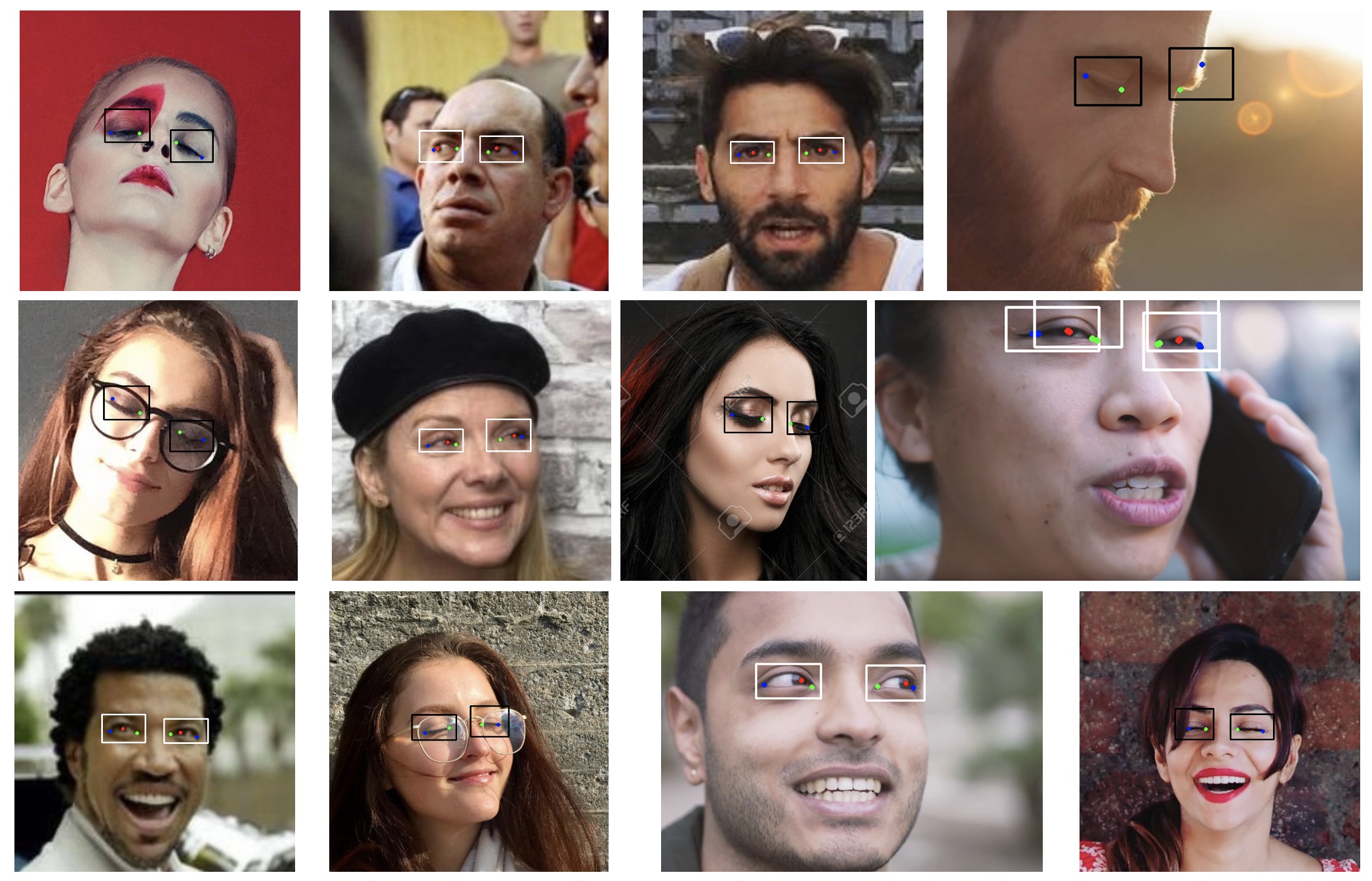}
  \caption{Sample visualization of predictions on the testing set of
    MIT Pupil Dataset. Last column shows some failure cases.}
  \label{fig:mit}
\end{figure}

We provide example visualizations of predictions of the proposed model on the three
testing datasets: MPIIGaze (Fig.~\ref{fig:mpii}), RT-BENE (Fig.~\ref{fig:rt}), and MIT Pupil Dataset testing set (Fig.~\ref{fig:mit}). The color of
bounding boxes indicates the eye state (white for open and black for closed).
The eye landmarks are visualized as color dots (blue for lateral canthus, green
for medial canthus, and red for pupil).

\section{Discussion}

Vision-based characterization of human attention allocation has been receiving increasing attention in recent HCI research, especially modeling related to eye dynamics, which shows great potential in real-world applications such as human-system engagement studies and applied driver monitoring. The main question we explore in this work is whether it's possible to develop a unified, end-to-end model for multiple eye-region analysis and eye-dynamics modeling tasks. Taking cognitive load estimation as an example, current approaches~\cite{fridman2017can, fridman2018cognitive} use either eye landmark positions or cropped eye images as input for an eye-dynamics modeling system, both of which require a high-accuracy eye and/or eye landmark detector to be run beforehand.

In our approach, instead of modeling eye dynamics through cropped eye images, we employ a shared deep neural network to extract image features. These are then used with different network heads to perform multiple eye-related tasks, including both low-level tasks like landmark detection and high-level tasks like cognitive load estimation. Multiple experiments show that by using a unified model, we can perform all tasks at almost no additional computational cost compared to a standard eye tracker, while also outperforming prior task-specific models in all tasks, including eye landmark detection, blink detection, and cognitive load estimation.

In the broader context of HCI research, we hope our work will inspire further investigations into more unified modeling of HCI and human factors tasks, rather than focusing solely on specific individual tasks. Our work demonstrates that by utilizing advanced deep learning techniques, the joint modeling of different yet correlated tasks can not only reduce computational cost but also improve the performance of each task. Our proposed CLERA model can support a multitude of HCI research activities involving human attention monitoring. The richness of its output and its capability for real-time monitoring can aid applications ranging from theoretical investigations of human attentional characteristics under various conditions of cognitive load or other states, to assessing the quality of engagement with different human-machine interface conceptual designs and actual implementations~\cite{joseph2020potential}. Furthermore, it can increase the practicality and relative cost-effectiveness of operator monitoring systems in aviation~\cite{ziv2016gaze}, air traffic control~\cite{ahlstrom2010eye,yazgan2021overview}, and power plant systems~\cite{yan2017effect,zhang2020study}. It can also address the increasing safety needs of drivers as they shift from primarily active driving to roles involving more system monitoring~\cite{hoedemaeker2007attuning,mehler2020is,pillai2022eye,ramakrishnan2021cognitive}. In summary, the CLERA model can facilitate the development of adaptive systems which account for variations in cognitive load, thus enhancing the viability of real-time operator support and fostering the improvement of human-centered systems. The large-scale dataset proposed in this work also offers a new source and benchmark for eye-region analysis, a need that has been highlighted in the literature~\cite{lan2022eyesyn}, and can be employed to support other research problems related to unified modeling.

\section{Limitations}

Previous research on cognitive load estimation has primarily utilized synthetic or controlled environments, such as driving simulators~\cite{liang2007real,pedrotti2014automatic}, tele-surgical robotic simulations~\cite{wu2020eye}, and simulation games~\cite{appel2021cross}. Although these environments offer a controlled and safe way to experiment with cognitive load, their results may not always be generalizable to real-world situations. Our work addresses this issue by focusing on the estimation of cognitive load in real-world settings, which are more complex and variable than controlled environments.

Our study has several limitations that should be considered. Firstly, our dataset for cognitive load estimation includes varying lighting conditions and camera placements, which may impact the accuracy of the methods used. We implemented two comparison methods from \cite{liang2007real,fridman2018cognitive} and found that both experienced a decrease in accuracy from above 80\% to around 60\% for the cognitive load level classification task. While our proposed method for cognitive load estimation shows a significant improvement over previous work, it may not be appropriate to use our model directly in its current form for real-world HCI applications due to performance and safety concerns. Additionally, our study focuses solely on a specific type of cognitive load estimation method that uses computer vision models with eye-dynamics input and does not explore other possible approaches such as glance detection and physiological signals like heart rates. However, we believe that our method could work well in simple and controlled environments without specific tuning or modifications, given its superior performance in our more challenging testing circumstances.

Nonetheless, our work provides valuable insights into the challenges of estimating cognitive load in naturalistic environments and aims to inform the development of more robust models that can be applied in practical settings. Specifically, the MIT Pupil Dataset proposed in this work could assist in the development of more accurate and robust models for eye-related analysis tasks, given that it is the largest open-source dataset in the field. The method proposed in this work can be extended to other HCI tasks such as facial analysis and emotion estimation and can be improved by using better deep learning architectures from the latest computer vision research.

\section{Conclusion}

In this work, we propose CLERA - a deep learning framework for joint cognitive load and eye region analysis. By using a detection model with two novel techniques: Localized Feature Tracking and Mask-Localized Regressor, the proposed model is capable of learning visual feature representations for precise eye bounding box and landmark detection. Additionally, it can track these representations over time and apply temporal modeling for cognitive load estimation. We also introduce the MIT Pupil Dataset, a large-scale, open-source dataset comprised of around 30k images of human faces with joint pupil, eye-openness, and landmark annotations.

The main contribution of our work lies in our demonstration that the tasks of eye-region analysis and eye-dynamics modeling can be jointly modeled. This approach ensures that the computational cost is on par with that of a common eye tracker. Moreover, our model is capable of outperforming prior work in all evaluated tasks, including cognitive load estimation, eye landmark detection, and blink estimation.

In terms of future work, we look forward to exploring other tasks in the area of human factors and human-centered computing that can be modeled through eye and facial movements using the proposed framework. This work also provides a new benchmark for eye-region analysis and can be utilized to support related research areas.

\begin{acks}
  The dataset used in this work is from work supported by {\em Veoneer}. The views and conclusions expressed are those of the authors and have not been sponsored, approved, or endorsed by {\em Veoneer}.
\end{acks}

\bibliographystyle{ACM-Reference-Format}
\bibliography{pupil}


\begin{thebibliography}{80}


\ifx \showCODEN    \undefined \def \showCODEN     #1{\unskip}     \fi
\ifx \showDOI      \undefined \def \showDOI       #1{#1}\fi
\ifx \showISBNx    \undefined \def \showISBNx     #1{\unskip}     \fi
\ifx \showISBNxiii \undefined \def \showISBNxiii  #1{\unskip}     \fi
\ifx \showISSN     \undefined \def \showISSN      #1{\unskip}     \fi
\ifx \showLCCN     \undefined \def \showLCCN      #1{\unskip}     \fi
\ifx \shownote     \undefined \def \shownote      #1{#1}          \fi
\ifx \showarticletitle \undefined \def \showarticletitle #1{#1}   \fi
\ifx \showURL      \undefined \def \showURL       {\relax}        \fi
\providecommand\bibfield[2]{#2}
\providecommand\bibinfo[2]{#2}
\providecommand\natexlab[1]{#1}
\providecommand\showeprint[2][]{arXiv:#2}

\bibitem[Ahlstrom(2010)]%
        {ahlstrom2010eye}
\bibfield{author}{\bibinfo{person}{Ulf Ahlstrom}.}
  \bibinfo{year}{2010}\natexlab{}.
\newblock \showarticletitle{An eye for the air traffic controller workload}. In
  \bibinfo{booktitle}{\emph{Journal of the Transportation Research Forum}},
  Vol.~\bibinfo{volume}{46}.
\newblock


\bibitem[Anas et~al\mbox{.}(2017)]%
        {anas2017online}
\bibfield{author}{\bibinfo{person}{Essa~R Anas}, \bibinfo{person}{Pedro
  Henriquez}, {and} \bibinfo{person}{Bogdan~J Matuszewski}.}
  \bibinfo{year}{2017}\natexlab{}.
\newblock \showarticletitle{Online Eye Status Detection in the Wild with
  Convolutional Neural Networks.}. In \bibinfo{booktitle}{\emph{VISIGRAPP (6:
  VISAPP)}}. \bibinfo{pages}{88--95}.
\newblock


\bibitem[Appel et~al\mbox{.}(2023)]%
        {appel2021cross}
\bibfield{author}{\bibinfo{person}{Tobias Appel}, \bibinfo{person}{Peter
  Gerjets}, \bibinfo{person}{Stefan Hoffmann}, \bibinfo{person}{Korbinian
  Moeller}, \bibinfo{person}{Manuel Ninaus}, \bibinfo{person}{Christian
  Scharinger}, \bibinfo{person}{Natalia Sevcenko}, \bibinfo{person}{Franz
  Wortha}, {and} \bibinfo{person}{Enkelejda Kasneci}.}
  \bibinfo{year}{2023}\natexlab{}.
\newblock \showarticletitle{Cross-Task and Cross-Participant Classification of
  Cognitive Load in an Emergency Simulation Game}.
\newblock \bibinfo{journal}{\emph{IEEE Transactions on Affective Computing}}
  \bibinfo{volume}{14}, \bibinfo{number}{2} (\bibinfo{year}{2023}),
  \bibinfo{pages}{1558--1571}.
\newblock
\urldef\tempurl%
\url{https://doi.org/10.1109/TAFFC.2021.3098237}
\showDOI{\tempurl}


\bibitem[Aracena et~al\mbox{.}(2015)]%
        {aracena2015neural}
\bibfield{author}{\bibinfo{person}{Claudio Aracena},
  \bibinfo{person}{Sebasti{\'a}n Basterrech}, \bibinfo{person}{V{\'a}clav
  Sn{\'a}el}, {and} \bibinfo{person}{Juan Vel{\'a}squez}.}
  \bibinfo{year}{2015}\natexlab{}.
\newblock \showarticletitle{Neural networks for emotion recognition based on
  eye tracking data}. In \bibinfo{booktitle}{\emph{2015 IEEE International
  Conference on Systems, Man, and Cybernetics}}. IEEE,
  \bibinfo{pages}{2632--2637}.
\newblock


\bibitem[Borji and Itti(2012)]%
        {borji2012state}
\bibfield{author}{\bibinfo{person}{Ali Borji} {and} \bibinfo{person}{Laurent
  Itti}.} \bibinfo{year}{2012}\natexlab{}.
\newblock \showarticletitle{State-of-the-art in visual attention modeling}.
\newblock \bibinfo{journal}{\emph{IEEE transactions on pattern analysis and
  machine intelligence}} \bibinfo{volume}{35}, \bibinfo{number}{1}
  (\bibinfo{year}{2012}), \bibinfo{pages}{185--207}.
\newblock


\bibitem[Cao et~al\mbox{.}(2017)]%
        {cao2017realtime}
\bibfield{author}{\bibinfo{person}{Zhe Cao}, \bibinfo{person}{Tomas Simon},
  \bibinfo{person}{Shih-En Wei}, {and} \bibinfo{person}{Yaser Sheikh}.}
  \bibinfo{year}{2017}\natexlab{}.
\newblock \showarticletitle{Realtime Multi-Person 2D Pose Estimation using Part
  Affinity Fields}. In \bibinfo{booktitle}{\emph{CVPR}}.
\newblock


\bibitem[Chen et~al\mbox{.}(2015)]%
        {chen2015real}
\bibfield{author}{\bibinfo{person}{Bo-Chun Chen}, \bibinfo{person}{Po-Chen Wu},
  {and} \bibinfo{person}{Shao-Yi Chien}.} \bibinfo{year}{2015}\natexlab{}.
\newblock \showarticletitle{Real-time eye localization, blink detection, and
  gaze estimation system without infrared illumination}. In
  \bibinfo{booktitle}{\emph{2015 IEEE International Conference on Image
  Processing (ICIP)}}. IEEE, \bibinfo{pages}{715--719}.
\newblock


\bibitem[Chen et~al\mbox{.}(2018)]%
        {chen2018cascaded}
\bibfield{author}{\bibinfo{person}{Yilun Chen}, \bibinfo{person}{Zhicheng
  Wang}, \bibinfo{person}{Yuxiang Peng}, \bibinfo{person}{Zhiqiang Zhang},
  \bibinfo{person}{Gang Yu}, {and} \bibinfo{person}{Jian Sun}.}
  \bibinfo{year}{2018}\natexlab{}.
\newblock \showarticletitle{Cascaded pyramid network for multi-person pose
  estimation}. In \bibinfo{booktitle}{\emph{Proceedings of the IEEE conference
  on computer vision and pattern recognition}}. \bibinfo{pages}{7103--7112}.
\newblock


\bibitem[Clay et~al\mbox{.}(2019)]%
        {clay2019eye}
\bibfield{author}{\bibinfo{person}{Viviane Clay}, \bibinfo{person}{Peter
  K{\"o}nig}, {and} \bibinfo{person}{Sabine Koenig}.}
  \bibinfo{year}{2019}\natexlab{}.
\newblock \showarticletitle{Eye tracking in virtual reality}.
\newblock \bibinfo{journal}{\emph{Journal of eye movement research}}
  \bibinfo{volume}{12}, \bibinfo{number}{1} (\bibinfo{year}{2019}).
\newblock


\bibitem[Cortacero et~al\mbox{.}(2019)]%
        {cortacero2019rt}
\bibfield{author}{\bibinfo{person}{K{\'e}vin Cortacero},
  \bibinfo{person}{Tobias Fischer}, {and} \bibinfo{person}{Yiannis Demiris}.}
  \bibinfo{year}{2019}\natexlab{}.
\newblock \showarticletitle{RT-BENE: A Dataset and Baselines for Real-Time
  Blink Estimation in Natural Environments}. In
  \bibinfo{booktitle}{\emph{Proceedings of the IEEE International Conference on
  Computer Vision Workshops}}. \bibinfo{pages}{0--0}.
\newblock


\bibitem[Cowie et~al\mbox{.}(2001)]%
        {cowie2001emotion}
\bibfield{author}{\bibinfo{person}{Roddy Cowie}, \bibinfo{person}{Ellen
  Douglas-Cowie}, \bibinfo{person}{Nicolas Tsapatsoulis},
  \bibinfo{person}{George Votsis}, \bibinfo{person}{Stefanos Kollias},
  \bibinfo{person}{Winfried Fellenz}, {and} \bibinfo{person}{John~G Taylor}.}
  \bibinfo{year}{2001}\natexlab{}.
\newblock \showarticletitle{Emotion recognition in human-computer interaction}.
\newblock \bibinfo{journal}{\emph{IEEE Signal processing magazine}}
  \bibinfo{volume}{18}, \bibinfo{number}{1} (\bibinfo{year}{2001}),
  \bibinfo{pages}{32--80}.
\newblock


\bibitem[Ding and Fridman(2019)]%
        {ding2019object}
\bibfield{author}{\bibinfo{person}{Li Ding} {and} \bibinfo{person}{Lex
  Fridman}.} \bibinfo{year}{2019}\natexlab{}.
\newblock \showarticletitle{Object as distribution}.
\newblock \bibinfo{journal}{\emph{arXiv preprint arXiv:1907.12929}}
  (\bibinfo{year}{2019}).
\newblock


\bibitem[Ding et~al\mbox{.}(2020a)]%
        {ding2020avt}
\bibfield{author}{\bibinfo{person}{Li Ding}, \bibinfo{person}{Michael Glazer},
  \bibinfo{person}{Meng Wang}, \bibinfo{person}{Bruce Mehler},
  \bibinfo{person}{Bryan Reimer}, {and} \bibinfo{person}{Lex Fridman}.}
  \bibinfo{year}{2020}\natexlab{a}.
\newblock \showarticletitle{Mit-avt clustered driving scene dataset: Evaluating
  perception systems in real-world naturalistic driving scenarios}. In
  \bibinfo{booktitle}{\emph{2020 IEEE Intelligent Vehicles Symposium (IV)}}.
  IEEE, \bibinfo{pages}{232--237}.
\newblock


\bibitem[Ding et~al\mbox{.}(2021a)]%
        {ding2021perceptual}
\bibfield{author}{\bibinfo{person}{Li Ding}, \bibinfo{person}{Rini Sherony},
  \bibinfo{person}{Bruce Mehler}, {and} \bibinfo{person}{Bryan Reimer}.}
  \bibinfo{year}{2021}\natexlab{a}.
\newblock \showarticletitle{Perceptual Evaluation of Driving Scene
  Segmentation}. In \bibinfo{booktitle}{\emph{2021 IEEE Intelligent Vehicles
  Symposium (IV)}}. IEEE, \bibinfo{pages}{1444--1450}.
\newblock


\bibitem[Ding et~al\mbox{.}(2020b)]%
        {ding2020drivesegm}
\bibfield{author}{\bibinfo{person}{Li Ding}, \bibinfo{person}{Jack
  Terwilliger}, \bibinfo{person}{Rini Sherony}, \bibinfo{person}{Bryan Reimer},
  {and} \bibinfo{person}{Lex Fridman}.} \bibinfo{year}{2020}\natexlab{b}.
\newblock \showarticletitle{MIT DriveSeg (Manual) Dataset for Dynamic Driving
  Scene Segmentation}.
\newblock \bibinfo{journal}{\emph{Massachusetts Institute of Technology AgeLab
  Technical Report 2020-1, Cambridge, MA}} (\bibinfo{year}{2020}).
\newblock


\bibitem[Ding et~al\mbox{.}(2020c)]%
        {ding2020drivesegs}
\bibfield{author}{\bibinfo{person}{Li Ding}, \bibinfo{person}{Jack
  Terwilliger}, \bibinfo{person}{Rini Sherony}, \bibinfo{person}{Bryan Reimer},
  {and} \bibinfo{person}{Lex Fridman}.} \bibinfo{year}{2020}\natexlab{c}.
\newblock \showarticletitle{MIT DriveSeg (Semi-auto) Dataset: Large-scale
  Semi-automated Annotation of Semantic Driving Scenes}.
\newblock \bibinfo{journal}{\emph{Massachusetts Institute of Technology AgeLab
  Technical Report 2020-2, Cambridge, MA}} (\bibinfo{year}{2020}).
\newblock


\bibitem[Ding et~al\mbox{.}(2021b)]%
        {ding2021value}
\bibfield{author}{\bibinfo{person}{Li Ding}, \bibinfo{person}{Jack
  Terwilliger}, \bibinfo{person}{Rini Sherony}, \bibinfo{person}{Bryan Reimer},
  {and} \bibinfo{person}{Lex Fridman}.} \bibinfo{year}{2021}\natexlab{b}.
\newblock \showarticletitle{Value of temporal dynamics information in driving
  scene segmentation}.
\newblock \bibinfo{journal}{\emph{IEEE Transactions on Intelligent Vehicles}}
  \bibinfo{volume}{7}, \bibinfo{number}{1} (\bibinfo{year}{2021}),
  \bibinfo{pages}{113--122}.
\newblock


\bibitem[Drutarovsky and Fogelton(2014)]%
        {drutarovsky2014eye}
\bibfield{author}{\bibinfo{person}{Tomas Drutarovsky} {and}
  \bibinfo{person}{Andrej Fogelton}.} \bibinfo{year}{2014}\natexlab{}.
\newblock \showarticletitle{Eye blink detection using variance of motion
  vectors}. In \bibinfo{booktitle}{\emph{European Conference on Computer
  Vision}}. Springer, \bibinfo{pages}{436--448}.
\newblock


\bibitem[Fischer(2001)]%
        {fischer2001user}
\bibfield{author}{\bibinfo{person}{Gerhard Fischer}.}
  \bibinfo{year}{2001}\natexlab{}.
\newblock \showarticletitle{User modeling in human--computer interaction}.
\newblock \bibinfo{journal}{\emph{User modeling and user-adapted interaction}}
  \bibinfo{volume}{11}, \bibinfo{number}{1} (\bibinfo{year}{2001}),
  \bibinfo{pages}{65--86}.
\newblock


\bibitem[Fischer et~al\mbox{.}(2018)]%
        {fischer2018rt}
\bibfield{author}{\bibinfo{person}{Tobias Fischer}, \bibinfo{person}{Hyung
  Jin~Chang}, {and} \bibinfo{person}{Yiannis Demiris}.}
  \bibinfo{year}{2018}\natexlab{}.
\newblock \showarticletitle{Rt-gene: Real-time eye gaze estimation in natural
  environments}. In \bibinfo{booktitle}{\emph{Proceedings of the European
  Conference on Computer Vision (ECCV)}}. \bibinfo{pages}{334--352}.
\newblock


\bibitem[Fogelton and Benesova(2016)]%
        {fogelton2016eye}
\bibfield{author}{\bibinfo{person}{A. Fogelton} {and} \bibinfo{person}{W.
  Benesova}.} \bibinfo{year}{2016}\natexlab{}.
\newblock \showarticletitle{Eye Blink Detection Based on Motion Vectors
  Analysis}.
\newblock \bibinfo{journal}{\emph{Comput. Vis. Image Underst.}}
  \bibinfo{volume}{148}, \bibinfo{number}{C} (\bibinfo{date}{jul}
  \bibinfo{year}{2016}), \bibinfo{pages}{23–33}.
\newblock
\showISSN{1077-3142}
\urldef\tempurl%
\url{https://doi.org/10.1016/j.cviu.2016.03.011}
\showDOI{\tempurl}


\bibitem[Fridman et~al\mbox{.}(2019a)]%
        {fridman2019advanced}
\bibfield{author}{\bibinfo{person}{Lex Fridman}, \bibinfo{person}{Daniel~E.
  Brown}, \bibinfo{person}{Michael Glazer}, \bibinfo{person}{William Angell},
  \bibinfo{person}{Spencer Dodd}, \bibinfo{person}{Benedikt Jenik},
  \bibinfo{person}{Jack Terwilliger}, \bibinfo{person}{Aleksandr Patsekin},
  \bibinfo{person}{Julia Kindelsberger}, \bibinfo{person}{Li Ding},
  \bibinfo{person}{Sean Seaman}, \bibinfo{person}{Alea Mehler},
  \bibinfo{person}{Andrew Sipperley}, \bibinfo{person}{Anthony Pettinato},
  \bibinfo{person}{Bobbie~D. Seppelt}, \bibinfo{person}{Linda Angell},
  \bibinfo{person}{Bruce Mehler}, {and} \bibinfo{person}{Bryan Reimer}.}
  \bibinfo{year}{2019}\natexlab{a}.
\newblock \showarticletitle{MIT Advanced Vehicle Technology Study: Large-Scale
  Naturalistic Driving Study of Driver Behavior and Interaction With
  Automation}.
\newblock \bibinfo{journal}{\emph{IEEE Access}}  \bibinfo{volume}{7}
  (\bibinfo{year}{2019}), \bibinfo{pages}{102021--102038}.
\newblock
\urldef\tempurl%
\url{https://doi.org/10.1109/ACCESS.2019.2926040}
\showDOI{\tempurl}


\bibitem[Fridman et~al\mbox{.}(2019b)]%
        {fridman2019arguing}
\bibfield{author}{\bibinfo{person}{Lex Fridman}, \bibinfo{person}{Li Ding},
  \bibinfo{person}{Benedikt Jenik}, {and} \bibinfo{person}{Bryan Reimer}.}
  \bibinfo{year}{2019}\natexlab{b}.
\newblock \showarticletitle{Arguing machines: Human supervision of black box AI
  systems that make life-critical decisions}. In
  \bibinfo{booktitle}{\emph{Proceedings of the IEEE/CVF Conference on Computer
  Vision and Pattern Recognition Workshops}}. \bibinfo{pages}{0--0}.
\newblock


\bibitem[Fridman et~al\mbox{.}(2018)]%
        {fridman2018cognitive}
\bibfield{author}{\bibinfo{person}{Lex Fridman}, \bibinfo{person}{Bryan
  Reimer}, \bibinfo{person}{Bruce Mehler}, {and} \bibinfo{person}{William~T
  Freeman}.} \bibinfo{year}{2018}\natexlab{}.
\newblock \showarticletitle{Cognitive load estimation in the wild}. In
  \bibinfo{booktitle}{\emph{Proceedings of the 2018 CHI Conference on Human
  Factors in Computing Systems}}. \bibinfo{pages}{1--9}.
\newblock


\bibitem[Fridman et~al\mbox{.}(2017)]%
        {fridman2017can}
\bibfield{author}{\bibinfo{person}{Lex Fridman}, \bibinfo{person}{Heishiro
  Toyoda}, \bibinfo{person}{Sean Seaman}, \bibinfo{person}{Bobbie Seppelt},
  \bibinfo{person}{Linda Angell}, \bibinfo{person}{Joonbum Lee},
  \bibinfo{person}{Bruce Mehler}, {and} \bibinfo{person}{Bryan Reimer}.}
  \bibinfo{year}{2017}\natexlab{}.
\newblock \showarticletitle{What can be predicted from six seconds of driver
  glances?}. In \bibinfo{booktitle}{\emph{Proceedings of the 2017 CHI
  Conference on Human Factors in Computing Systems}}.
  \bibinfo{pages}{2805--2813}.
\newblock


\bibitem[Fuhl et~al\mbox{.}(2015)]%
        {fuhl2015excuse}
\bibfield{author}{\bibinfo{person}{Wolfgang Fuhl}, \bibinfo{person}{Thomas
  K{\"u}bler}, \bibinfo{person}{Katrin Sippel}, \bibinfo{person}{Wolfgang
  Rosenstiel}, {and} \bibinfo{person}{Enkelejda Kasneci}.}
  \bibinfo{year}{2015}\natexlab{}.
\newblock \showarticletitle{Excuse: Robust pupil detection in real-world
  scenarios}. In \bibinfo{booktitle}{\emph{International Conference on Computer
  Analysis of Images and Patterns}}. Springer, \bibinfo{pages}{39--51}.
\newblock


\bibitem[Gao et~al\mbox{.}(2007)]%
        {gao2007cas}
\bibfield{author}{\bibinfo{person}{Wen Gao}, \bibinfo{person}{Bo Cao},
  \bibinfo{person}{Shiguang Shan}, \bibinfo{person}{Xilin Chen},
  \bibinfo{person}{Delong Zhou}, \bibinfo{person}{Xiaohua Zhang}, {and}
  \bibinfo{person}{Debin Zhao}.} \bibinfo{year}{2007}\natexlab{}.
\newblock \showarticletitle{The CAS-PEAL large-scale Chinese face database and
  baseline evaluations}.
\newblock \bibinfo{journal}{\emph{IEEE Transactions on Systems, Man, and
  Cybernetics-Part A: Systems and Humans}} \bibinfo{volume}{38},
  \bibinfo{number}{1} (\bibinfo{year}{2007}), \bibinfo{pages}{149--161}.
\newblock


\bibitem[Garbin et~al\mbox{.}(2019)]%
        {garbin2019openeds}
\bibfield{author}{\bibinfo{person}{Stephan~J Garbin}, \bibinfo{person}{Yiru
  Shen}, \bibinfo{person}{Immo Schuetz}, \bibinfo{person}{Robert Cavin},
  \bibinfo{person}{Gregory Hughes}, {and} \bibinfo{person}{Sachin~S Talathi}.}
  \bibinfo{year}{2019}\natexlab{}.
\newblock \showarticletitle{Openeds: Open eye dataset}.
\newblock \bibinfo{journal}{\emph{arXiv preprint arXiv:1905.03702}}
  (\bibinfo{year}{2019}).
\newblock


\bibitem[Goldberg and Wichansky(2003)]%
        {goldberg2003eye}
\bibfield{author}{\bibinfo{person}{Joseph~H Goldberg} {and}
  \bibinfo{person}{Anna~M Wichansky}.} \bibinfo{year}{2003}\natexlab{}.
\newblock \showarticletitle{Eye tracking in usability evaluation: A
  practitioner's guide}.
\newblock In \bibinfo{booktitle}{\emph{the Mind's Eye}}.
  \bibinfo{publisher}{Elsevier}, \bibinfo{pages}{493--516}.
\newblock


\bibitem[Haapalainen et~al\mbox{.}(2010)]%
        {haapalainen2010psycho}
\bibfield{author}{\bibinfo{person}{Eija Haapalainen}, \bibinfo{person}{SeungJun
  Kim}, \bibinfo{person}{Jodi~F Forlizzi}, {and} \bibinfo{person}{Anind~K
  Dey}.} \bibinfo{year}{2010}\natexlab{}.
\newblock \showarticletitle{Psycho-physiological measures for assessing
  cognitive load}. In \bibinfo{booktitle}{\emph{Proceedings of the 12th ACM
  international conference on Ubiquitous computing}}. ACM,
  \bibinfo{pages}{301--310}.
\newblock


\bibitem[Harezlak and Kasprowski(2018)]%
        {harezlak2018application}
\bibfield{author}{\bibinfo{person}{Katarzyna Harezlak} {and}
  \bibinfo{person}{Pawel Kasprowski}.} \bibinfo{year}{2018}\natexlab{}.
\newblock \showarticletitle{Application of eye tracking in medicine: A survey,
  research issues and challenges}.
\newblock \bibinfo{journal}{\emph{Computerized Medical Imaging and Graphics}}
  \bibinfo{volume}{65} (\bibinfo{year}{2018}), \bibinfo{pages}{176--190}.
\newblock
\showISSN{0895-6111}
\urldef\tempurl%
\url{https://doi.org/10.1016/j.compmedimag.2017.04.006}
\showDOI{\tempurl}
\newblock
\shownote{Advances in Biomedical Image Processing}.


\bibitem[He et~al\mbox{.}(2017)]%
        {he2017mask}
\bibfield{author}{\bibinfo{person}{Kaiming He}, \bibinfo{person}{Georgia
  Gkioxari}, \bibinfo{person}{Piotr Doll{\'a}r}, {and} \bibinfo{person}{Ross
  Girshick}.} \bibinfo{year}{2017}\natexlab{}.
\newblock \showarticletitle{Mask r-cnn}. In \bibinfo{booktitle}{\emph{Computer
  Vision (ICCV), 2017 IEEE International Conference on}}. IEEE,
  \bibinfo{pages}{2980--2988}.
\newblock


\bibitem[He et~al\mbox{.}(2016)]%
        {he2016deep}
\bibfield{author}{\bibinfo{person}{Kaiming He}, \bibinfo{person}{Xiangyu
  Zhang}, \bibinfo{person}{Shaoqing Ren}, {and} \bibinfo{person}{Jian Sun}.}
  \bibinfo{year}{2016}\natexlab{}.
\newblock \showarticletitle{Deep residual learning for image recognition}. In
  \bibinfo{booktitle}{\emph{Proceedings of the IEEE conference on computer
  vision and pattern recognition}}. \bibinfo{pages}{770--778}.
\newblock


\bibitem[Hoedemaeker and Neerincx(2007)]%
        {hoedemaeker2007attuning}
\bibfield{author}{\bibinfo{person}{Marika Hoedemaeker} {and}
  \bibinfo{person}{Mark Neerincx}.} \bibinfo{year}{2007}\natexlab{}.
\newblock \showarticletitle{Attuning in-car user interfaces to the momentary
  cognitive load}. In \bibinfo{booktitle}{\emph{International Conference on
  Foundations of Augmented Cognition}}. Springer, \bibinfo{pages}{286--293}.
\newblock


\bibitem[Huang et~al\mbox{.}(2008)]%
        {huang2008labeled}
\bibfield{author}{\bibinfo{person}{Gary~B Huang}, \bibinfo{person}{Marwan
  Mattar}, \bibinfo{person}{Tamara Berg}, {and} \bibinfo{person}{Eric
  Learned-Miller}.} \bibinfo{year}{2008}\natexlab{}.
\newblock \showarticletitle{Labeled faces in the wild: A database forstudying
  face recognition in unconstrained environments}. In
  \bibinfo{booktitle}{\emph{Workshop on faces in'Real-Life'Images: detection,
  alignment, and recognition}}.
\newblock


\bibitem[Ioffe and Szegedy(2015)]%
        {ioffe2015batch}
\bibfield{author}{\bibinfo{person}{Sergey Ioffe} {and}
  \bibinfo{person}{Christian Szegedy}.} \bibinfo{year}{2015}\natexlab{}.
\newblock \showarticletitle{Batch normalization: Accelerating deep network
  training by reducing internal covariate shift}. In
  \bibinfo{booktitle}{\emph{International conference on machine learning}}.
  PMLR, \bibinfo{pages}{448--456}.
\newblock


\bibitem[Joseph and Murugesh(2020)]%
        {joseph2020potential}
\bibfield{author}{\bibinfo{person}{Antony~William Joseph} {and}
  \bibinfo{person}{Ramaswamy Murugesh}.} \bibinfo{year}{2020}\natexlab{}.
\newblock \showarticletitle{Potential eye tracking metrics and indicators to
  measure cognitive load in human-computer interaction research}.
\newblock \bibinfo{journal}{\emph{J. Sci. Res}} \bibinfo{volume}{64},
  \bibinfo{number}{1} (\bibinfo{year}{2020}), \bibinfo{pages}{168--175}.
\newblock


\bibitem[Joseph et~al\mbox{.}(2021)]%
        {joseph2021modeling}
\bibfield{author}{\bibinfo{person}{Antony~William Joseph},
  \bibinfo{person}{J~Sharmila Vaiz}, {and} \bibinfo{person}{Ramaswami
  Murugesh}.} \bibinfo{year}{2021}\natexlab{}.
\newblock \showarticletitle{Modeling Cognitive Load in Mobile Human Computer
  Interaction Using Eye Tracking Metrics}. In
  \bibinfo{booktitle}{\emph{International Conference on Applied Human Factors
  and Ergonomics}}. Springer, \bibinfo{pages}{99--106}.
\newblock


\bibitem[Kim et~al\mbox{.}(2019)]%
        {kim2019nvgaze}
\bibfield{author}{\bibinfo{person}{Joohwan Kim}, \bibinfo{person}{Michael
  Stengel}, \bibinfo{person}{Alexander Majercik}, \bibinfo{person}{Shalini
  De~Mello}, \bibinfo{person}{David Dunn}, \bibinfo{person}{Samuli Laine},
  \bibinfo{person}{Morgan McGuire}, {and} \bibinfo{person}{David Luebke}.}
  \bibinfo{year}{2019}\natexlab{}.
\newblock \showarticletitle{Nvgaze: An anatomically-informed dataset for
  low-latency, near-eye gaze estimation}. In
  \bibinfo{booktitle}{\emph{Proceedings of the 2019 CHI Conference on Human
  Factors in Computing Systems}}. \bibinfo{pages}{1--12}.
\newblock


\bibitem[Kim et~al\mbox{.}(2017)]%
        {kim2017study}
\bibfield{author}{\bibinfo{person}{Ki Kim}, \bibinfo{person}{Hyung Hong},
  \bibinfo{person}{Gi Nam}, {and} \bibinfo{person}{Kang Park}.}
  \bibinfo{year}{2017}\natexlab{}.
\newblock \showarticletitle{A study of deep CNN-based classification of open
  and closed eyes using a visible light camera sensor}.
\newblock \bibinfo{journal}{\emph{Sensors}} \bibinfo{volume}{17},
  \bibinfo{number}{7} (\bibinfo{year}{2017}), \bibinfo{pages}{1534}.
\newblock


\bibitem[Lalonde et~al\mbox{.}(2007)]%
        {lalonde2007real}
\bibfield{author}{\bibinfo{person}{Marc Lalonde}, \bibinfo{person}{David
  Byrns}, \bibinfo{person}{Langis Gagnon}, \bibinfo{person}{Normand Teasdale},
  {and} \bibinfo{person}{Denis Laurendeau}.} \bibinfo{year}{2007}\natexlab{}.
\newblock \showarticletitle{Real-time eye blink detection with GPU-based SIFT
  tracking}. In \bibinfo{booktitle}{\emph{Fourth Canadian Conference on
  Computer and Robot Vision (CRV'07)}}. IEEE, \bibinfo{pages}{481--487}.
\newblock


\bibitem[Lan et~al\mbox{.}(2022)]%
        {lan2022eyesyn}
\bibfield{author}{\bibinfo{person}{Guohao Lan}, \bibinfo{person}{Tim Scargill},
  {and} \bibinfo{person}{Maria Gorlatova}.} \bibinfo{year}{2022}\natexlab{}.
\newblock \showarticletitle{EyeSyn: Psychology-inspired Eye Movement Synthesis
  for Gaze-based Activity Recognition}. In
  \bibinfo{booktitle}{\emph{Proceedings of ACM/IEEE IPSN}}.
\newblock


\bibitem[L{\'e}v{\^e}que et~al\mbox{.}(2018)]%
        {leveque2018state}
\bibfield{author}{\bibinfo{person}{Lucie L{\'e}v{\^e}que},
  \bibinfo{person}{Hilde Bosmans}, \bibinfo{person}{Lesley Cockmartin}, {and}
  \bibinfo{person}{Hantao Liu}.} \bibinfo{year}{2018}\natexlab{}.
\newblock \showarticletitle{State of the Art: Eye-Tracking Studies in Medical
  Imaging}.
\newblock \bibinfo{journal}{\emph{IEEE Access}}  \bibinfo{volume}{6}
  (\bibinfo{year}{2018}), \bibinfo{pages}{37023--37034}.
\newblock
\urldef\tempurl%
\url{https://doi.org/10.1109/ACCESS.2018.2851451}
\showDOI{\tempurl}


\bibitem[Liang et~al\mbox{.}(2007)]%
        {liang2007real}
\bibfield{author}{\bibinfo{person}{Yulan Liang}, \bibinfo{person}{Michelle~L
  Reyes}, {and} \bibinfo{person}{John~D Lee}.} \bibinfo{year}{2007}\natexlab{}.
\newblock \showarticletitle{Real-time detection of driver cognitive distraction
  using support vector machines}.
\newblock \bibinfo{journal}{\emph{IEEE transactions on intelligent
  transportation systems}} \bibinfo{volume}{8}, \bibinfo{number}{2}
  (\bibinfo{year}{2007}), \bibinfo{pages}{340--350}.
\newblock


\bibitem[Lim et~al\mbox{.}(2020)]%
        {lim2020emotion}
\bibfield{author}{\bibinfo{person}{Jia~Zheng Lim}, \bibinfo{person}{James
  Mountstephens}, {and} \bibinfo{person}{Jason Teo}.}
  \bibinfo{year}{2020}\natexlab{}.
\newblock \showarticletitle{Emotion recognition using eye-tracking: taxonomy,
  review and current challenges}.
\newblock \bibinfo{journal}{\emph{Sensors}} \bibinfo{volume}{20},
  \bibinfo{number}{8} (\bibinfo{year}{2020}), \bibinfo{pages}{2384}.
\newblock


\bibitem[Lin et~al\mbox{.}(2014)]%
        {lin2014microsoft}
\bibfield{author}{\bibinfo{person}{Tsung-Yi Lin}, \bibinfo{person}{Michael
  Maire}, \bibinfo{person}{Serge Belongie}, \bibinfo{person}{James Hays},
  \bibinfo{person}{Pietro Perona}, \bibinfo{person}{Deva Ramanan},
  \bibinfo{person}{Piotr Doll{\'a}r}, {and} \bibinfo{person}{C~Lawrence
  Zitnick}.} \bibinfo{year}{2014}\natexlab{}.
\newblock \showarticletitle{Microsoft coco: Common objects in context}. In
  \bibinfo{booktitle}{\emph{European conference on computer vision}}. Springer,
  \bibinfo{pages}{740--755}.
\newblock


\bibitem[Lu et~al\mbox{.}(2014)]%
        {lu2014adaptive}
\bibfield{author}{\bibinfo{person}{Feng Lu}, \bibinfo{person}{Yusuke Sugano},
  \bibinfo{person}{Takahiro Okabe}, {and} \bibinfo{person}{Yoichi Sato}.}
  \bibinfo{year}{2014}\natexlab{}.
\newblock \showarticletitle{Adaptive linear regression for appearance-based
  gaze estimation}.
\newblock \bibinfo{journal}{\emph{IEEE transactions on pattern analysis and
  machine intelligence}} \bibinfo{volume}{36}, \bibinfo{number}{10}
  (\bibinfo{year}{2014}), \bibinfo{pages}{2033--2046}.
\newblock


\bibitem[Majaranta and Bulling(2014)]%
        {majaranta2014eye}
\bibfield{author}{\bibinfo{person}{P{\"a}ivi Majaranta} {and}
  \bibinfo{person}{Andreas Bulling}.} \bibinfo{year}{2014}\natexlab{}.
\newblock \showarticletitle{Eye tracking and eye-based human--computer
  interaction}.
\newblock In \bibinfo{booktitle}{\emph{Advances in physiological computing}}.
  \bibinfo{publisher}{Springer}, \bibinfo{pages}{39--65}.
\newblock


\bibitem[Mehler(2020)]%
        {mehler2020is}
\bibfield{author}{\bibinfo{person}{Bruce Mehler}.}
  \bibinfo{year}{2020}\natexlab{}.
\newblock \bibinfo{title}{Is supportive driver monitoring needed to maximize
  trust, use, and the safety-benefits of collaborative automation?}
\newblock
\newblock


\bibitem[Mehler et~al\mbox{.}(2012)]%
        {mehler2012sensitivity}
\bibfield{author}{\bibinfo{person}{Bruce Mehler}, \bibinfo{person}{Bryan
  Reimer}, {and} \bibinfo{person}{Joseph~F Coughlin}.}
  \bibinfo{year}{2012}\natexlab{}.
\newblock \showarticletitle{Sensitivity of physiological measures for detecting
  systematic variations in cognitive demand from a working memory task: an
  on-road study across three age groups}.
\newblock \bibinfo{journal}{\emph{Human factors}} \bibinfo{volume}{54},
  \bibinfo{number}{3} (\bibinfo{year}{2012}), \bibinfo{pages}{396--412}.
\newblock


\bibitem[Oliveira et~al\mbox{.}(2021)]%
        {oliveira2021computer}
\bibfield{author}{\bibinfo{person}{Jessica~S Oliveira},
  \bibinfo{person}{Felipe~O Franco}, \bibinfo{person}{Mirian~C Revers},
  \bibinfo{person}{Andr{\'e}ia~F Silva}, \bibinfo{person}{Joana Portolese},
  \bibinfo{person}{Helena Brentani}, \bibinfo{person}{Ariane Machado-Lima},
  {and} \bibinfo{person}{F{\'a}tima~LS Nunes}.}
  \bibinfo{year}{2021}\natexlab{}.
\newblock \showarticletitle{Computer-aided autism diagnosis based on visual
  attention models using eye tracking}.
\newblock \bibinfo{journal}{\emph{Scientific reports}} \bibinfo{volume}{11},
  \bibinfo{number}{1} (\bibinfo{year}{2021}), \bibinfo{pages}{1--11}.
\newblock


\bibitem[Paas et~al\mbox{.}(2003)]%
        {paas2003cognitive}
\bibfield{author}{\bibinfo{person}{Fred Paas}, \bibinfo{person}{Juhani~E
  Tuovinen}, \bibinfo{person}{Huib Tabbers}, {and} \bibinfo{person}{Pascal~WM
  Van~Gerven}.} \bibinfo{year}{2003}\natexlab{}.
\newblock \showarticletitle{Cognitive load measurement as a means to advance
  cognitive load theory}.
\newblock \bibinfo{journal}{\emph{Educational psychologist}}
  \bibinfo{volume}{38}, \bibinfo{number}{1} (\bibinfo{year}{2003}),
  \bibinfo{pages}{63--71}.
\newblock


\bibitem[Pan et~al\mbox{.}(2007)]%
        {pan2007eyeblink}
\bibfield{author}{\bibinfo{person}{Gang Pan}, \bibinfo{person}{Lin Sun},
  \bibinfo{person}{Zhaohui Wu}, {and} \bibinfo{person}{Shihong Lao}.}
  \bibinfo{year}{2007}\natexlab{}.
\newblock \showarticletitle{Eyeblink-based anti-spoofing in face recognition
  from a generic webcamera}. In \bibinfo{booktitle}{\emph{2007 IEEE 11th
  International Conference on Computer Vision}}. IEEE, \bibinfo{pages}{1--8}.
\newblock


\bibitem[Papandreou et~al\mbox{.}(2017)]%
        {papandreou2017towards}
\bibfield{author}{\bibinfo{person}{George Papandreou}, \bibinfo{person}{Tyler
  Zhu}, \bibinfo{person}{Nori Kanazawa}, \bibinfo{person}{Alexander Toshev},
  \bibinfo{person}{Jonathan Tompson}, \bibinfo{person}{Chris Bregler}, {and}
  \bibinfo{person}{Kevin Murphy}.} \bibinfo{year}{2017}\natexlab{}.
\newblock \showarticletitle{Towards accurate multi-person pose estimation in
  the wild}. In \bibinfo{booktitle}{\emph{Proceedings of the IEEE Conference on
  Computer Vision and Pattern Recognition}}. \bibinfo{pages}{4903--4911}.
\newblock


\bibitem[Patney et~al\mbox{.}(2016)]%
        {patney2016perceptually}
\bibfield{author}{\bibinfo{person}{Anjul Patney}, \bibinfo{person}{Joohwan
  Kim}, \bibinfo{person}{Marco Salvi}, \bibinfo{person}{Anton Kaplanyan},
  \bibinfo{person}{Chris Wyman}, \bibinfo{person}{Nir Benty},
  \bibinfo{person}{Aaron Lefohn}, {and} \bibinfo{person}{David Luebke}.}
  \bibinfo{year}{2016}\natexlab{}.
\newblock \showarticletitle{Perceptually-based foveated virtual reality}.
\newblock In \bibinfo{booktitle}{\emph{ACM SIGGRAPH 2016 Emerging
  Technologies}}. \bibinfo{pages}{1--2}.
\newblock


\bibitem[Pedrotti et~al\mbox{.}(2014)]%
        {pedrotti2014automatic}
\bibfield{author}{\bibinfo{person}{Marco Pedrotti},
  \bibinfo{person}{Mohammad~Ali Mirzaei}, \bibinfo{person}{Adrien Tedesco},
  \bibinfo{person}{Jean-R{\'e}my Chardonnet}, \bibinfo{person}{Fr{\'e}d{\'e}ric
  M{\'e}rienne}, \bibinfo{person}{Simone Benedetto}, {and}
  \bibinfo{person}{Thierry Baccino}.} \bibinfo{year}{2014}\natexlab{}.
\newblock \showarticletitle{Automatic stress classification with pupil diameter
  analysis}.
\newblock \bibinfo{journal}{\emph{International Journal of Human-Computer
  Interaction}} \bibinfo{volume}{30}, \bibinfo{number}{3}
  (\bibinfo{year}{2014}), \bibinfo{pages}{220--236}.
\newblock


\bibitem[Pillai et~al\mbox{.}(2022)]%
        {pillai2022eye}
\bibfield{author}{\bibinfo{person}{Prarthana Pillai},
  \bibinfo{person}{Balakumar Balasingam}, \bibinfo{person}{Yong~Hoon Kim},
  \bibinfo{person}{Chris Lee}, {and} \bibinfo{person}{Francesco Biondi}.}
  \bibinfo{year}{2022}\natexlab{}.
\newblock \showarticletitle{Eye-Gaze Metrics for Cognitive Load Detection on a
  Driving Simulator}.
\newblock \bibinfo{journal}{\emph{IEEE/ASME Transactions on Mechatronics}}
  \bibinfo{volume}{27}, \bibinfo{number}{4} (\bibinfo{year}{2022}),
  \bibinfo{pages}{2134--2141}.
\newblock
\urldef\tempurl%
\url{https://doi.org/10.1109/TMECH.2022.3175774}
\showDOI{\tempurl}


\bibitem[Ramakrishnan et~al\mbox{.}(2021)]%
        {ramakrishnan2021cognitive}
\bibfield{author}{\bibinfo{person}{P Ramakrishnan}, \bibinfo{person}{B
  Balasingam}, {and} \bibinfo{person}{F Biondi}.}
  \bibinfo{year}{2021}\natexlab{}.
\newblock \showarticletitle{Cognitive load estimation for adaptive
  human--machine system automation}.
\newblock In \bibinfo{booktitle}{\emph{Learning control}}.
  \bibinfo{publisher}{Elsevier}, \bibinfo{pages}{35--58}.
\newblock


\bibitem[Recarte and Nunes(2003)]%
        {recarte2003mental}
\bibfield{author}{\bibinfo{person}{Miguel~A Recarte} {and}
  \bibinfo{person}{Luis~M Nunes}.} \bibinfo{year}{2003}\natexlab{}.
\newblock \showarticletitle{Mental workload while driving: effects on visual
  search, discrimination, and decision making.}
\newblock \bibinfo{journal}{\emph{Journal of experimental psychology: Applied}}
  \bibinfo{volume}{9}, \bibinfo{number}{2} (\bibinfo{year}{2003}),
  \bibinfo{pages}{119}.
\newblock


\bibitem[Redmon et~al\mbox{.}(2016)]%
        {redmon2016you}
\bibfield{author}{\bibinfo{person}{Joseph Redmon}, \bibinfo{person}{Santosh
  Divvala}, \bibinfo{person}{Ross Girshick}, {and} \bibinfo{person}{Ali
  Farhadi}.} \bibinfo{year}{2016}\natexlab{}.
\newblock \showarticletitle{You only look once: Unified, real-time object
  detection}. In \bibinfo{booktitle}{\emph{Proceedings of the IEEE conference
  on computer vision and pattern recognition}}. \bibinfo{pages}{779--788}.
\newblock


\bibitem[Redmon and Farhadi(2018)]%
        {redmon2018yolov3}
\bibfield{author}{\bibinfo{person}{Joseph Redmon} {and} \bibinfo{person}{Ali
  Farhadi}.} \bibinfo{year}{2018}\natexlab{}.
\newblock \showarticletitle{Yolov3: An incremental improvement}.
\newblock \bibinfo{journal}{\emph{arXiv preprint arXiv:1804.02767}}
  (\bibinfo{year}{2018}).
\newblock


\bibitem[Reimer et~al\mbox{.}(2012)]%
        {reimer2012field}
\bibfield{author}{\bibinfo{person}{Bryan Reimer}, \bibinfo{person}{Bruce
  Mehler}, \bibinfo{person}{Ying Wang}, {and} \bibinfo{person}{Joseph~F
  Coughlin}.} \bibinfo{year}{2012}\natexlab{}.
\newblock \showarticletitle{A field study on the impact of variations in
  short-term memory demands on drivers' visual attention and driving
  performance across three age groups}.
\newblock \bibinfo{journal}{\emph{Human Factors}} \bibinfo{volume}{54},
  \bibinfo{number}{3} (\bibinfo{year}{2012}), \bibinfo{pages}{454--468}.
\newblock


\bibitem[Ren et~al\mbox{.}(2015)]%
        {ren2015faster}
\bibfield{author}{\bibinfo{person}{Shaoqing Ren}, \bibinfo{person}{Kaiming He},
  \bibinfo{person}{Ross Girshick}, {and} \bibinfo{person}{Jian Sun}.}
  \bibinfo{year}{2015}\natexlab{}.
\newblock \showarticletitle{Faster r-cnn: Towards real-time object detection
  with region proposal networks}. In \bibinfo{booktitle}{\emph{Advances in
  neural information processing systems}}. \bibinfo{pages}{91--99}.
\newblock


\bibitem[Schillingmann and Nagai(2015)]%
        {schillingmann2015yet}
\bibfield{author}{\bibinfo{person}{Lars Schillingmann} {and}
  \bibinfo{person}{Yukie Nagai}.} \bibinfo{year}{2015}\natexlab{}.
\newblock \showarticletitle{Yet another gaze detector: An embodied calibration
  free system for the iCub robot}. In \bibinfo{booktitle}{\emph{2015 IEEE-RAS
  15th International Conference on Humanoid Robots (Humanoids)}}. IEEE,
  \bibinfo{pages}{8--13}.
\newblock


\bibitem[Siegfried et~al\mbox{.}(2019)]%
        {siegfried2019deep}
\bibfield{author}{\bibinfo{person}{R{\'e}my Siegfried}, \bibinfo{person}{Yu
  Yu}, {and} \bibinfo{person}{Jean-Marc Odobez}.}
  \bibinfo{year}{2019}\natexlab{}.
\newblock \showarticletitle{A deep learning approach for robust head pose
  independent eye movements recognition from videos}. In
  \bibinfo{booktitle}{\emph{Proceedings of the 11th ACM Symposium on Eye
  Tracking Research \& Applications}}. ACM, \bibinfo{pages}{31}.
\newblock


\bibitem[Simonyan and Zisserman(2015)]%
        {simonyan2014very}
\bibfield{author}{\bibinfo{person}{Karen Simonyan} {and}
  \bibinfo{person}{Andrew Zisserman}.} \bibinfo{year}{2015}\natexlab{}.
\newblock \showarticletitle{Very Deep Convolutional Networks for Large-Scale
  Image Recognition}. In \bibinfo{booktitle}{\emph{International Conference on
  Learning Representations}}.
\newblock


\bibitem[Song et~al\mbox{.}(2014)]%
        {song2014eyes}
\bibfield{author}{\bibinfo{person}{Fengyi Song}, \bibinfo{person}{Xiaoyang
  Tan}, \bibinfo{person}{Xue Liu}, {and} \bibinfo{person}{Songcan Chen}.}
  \bibinfo{year}{2014}\natexlab{}.
\newblock \showarticletitle{Eyes closeness detection from still images with
  multi-scale histograms of principal oriented gradients}.
\newblock \bibinfo{journal}{\emph{Pattern Recognition}} \bibinfo{volume}{47},
  \bibinfo{number}{9} (\bibinfo{year}{2014}), \bibinfo{pages}{2825--2838}.
\newblock


\bibitem[Sweller et~al\mbox{.}(2011)]%
        {sweller2011measuring}
\bibfield{author}{\bibinfo{person}{John Sweller}, \bibinfo{person}{Paul Ayres},
  {and} \bibinfo{person}{Slava Kalyuga}.} \bibinfo{year}{2011}\natexlab{}.
\newblock \showarticletitle{Measuring cognitive load}.
\newblock In \bibinfo{booktitle}{\emph{Cognitive load theory}}.
  \bibinfo{publisher}{Springer}, \bibinfo{pages}{71--85}.
\newblock


\bibitem[{\'S}wirski et~al\mbox{.}(2012)]%
        {swirski2012robust}
\bibfield{author}{\bibinfo{person}{Lech {\'S}wirski}, \bibinfo{person}{Andreas
  Bulling}, {and} \bibinfo{person}{Neil Dodgson}.}
  \bibinfo{year}{2012}\natexlab{}.
\newblock \showarticletitle{Robust real-time pupil tracking in highly off-axis
  images}. In \bibinfo{booktitle}{\emph{Proceedings of the Symposium on Eye
  Tracking Research and Applications}}. \bibinfo{pages}{173--176}.
\newblock


\bibitem[Tonsen et~al\mbox{.}(2016)]%
        {tonsen2016labelled}
\bibfield{author}{\bibinfo{person}{Marc Tonsen}, \bibinfo{person}{Xucong
  Zhang}, \bibinfo{person}{Yusuke Sugano}, {and} \bibinfo{person}{Andreas
  Bulling}.} \bibinfo{year}{2016}\natexlab{}.
\newblock \showarticletitle{Labelled pupils in the wild: a dataset for studying
  pupil detection in unconstrained environments}. In
  \bibinfo{booktitle}{\emph{Proceedings of the Ninth Biennial ACM Symposium on
  Eye Tracking Research \& Applications}}. \bibinfo{pages}{139--142}.
\newblock


\bibitem[Wang et~al\mbox{.}(2014)]%
        {wang2014sensitivity}
\bibfield{author}{\bibinfo{person}{Ying Wang}, \bibinfo{person}{Bryan Reimer},
  \bibinfo{person}{Jonathan Dobres}, {and} \bibinfo{person}{Bruce Mehler}.}
  \bibinfo{year}{2014}\natexlab{}.
\newblock \showarticletitle{The sensitivity of different methodologies for
  characterizing drivers' gaze concentration under increased cognitive demand}.
\newblock \bibinfo{journal}{\emph{Transportation Research Part F: Traffic
  Psychology and Behaviour}}  \bibinfo{volume}{26} (\bibinfo{year}{2014}),
  \bibinfo{pages}{227--237}.
\newblock
\showISSN{1369-8478}
\urldef\tempurl%
\url{https://doi.org/10.1016/j.trf.2014.08.003}
\showDOI{\tempurl}


\bibitem[Weber(2022)]%
        {faces1999database}
\bibfield{author}{\bibinfo{person}{Markus Weber}.}
  \bibinfo{year}{2022}\natexlab{}.
\newblock \bibinfo{title}{Caltech Face Dataset 1999}.
\newblock
\newblock
\urldef\tempurl%
\url{https://doi.org/10.22002/D1.20237}
\showDOI{\tempurl}


\bibitem[Wu et~al\mbox{.}(2020)]%
        {wu2020eye}
\bibfield{author}{\bibinfo{person}{Chuhao Wu}, \bibinfo{person}{Jackie Cha},
  \bibinfo{person}{Jay Sulek}, \bibinfo{person}{Tian Zhou},
  \bibinfo{person}{Chandru~P Sundaram}, \bibinfo{person}{Juan Wachs}, {and}
  \bibinfo{person}{Denny Yu}.} \bibinfo{year}{2020}\natexlab{}.
\newblock \showarticletitle{Eye-tracking metrics predict perceived workload in
  robotic surgical skills training}.
\newblock \bibinfo{journal}{\emph{Human factors}} \bibinfo{volume}{62},
  \bibinfo{number}{8} (\bibinfo{year}{2020}), \bibinfo{pages}{1365--1386}.
\newblock


\bibitem[Yan et~al\mbox{.}(2017)]%
        {yan2017effect}
\bibfield{author}{\bibinfo{person}{Shengyuan Yan}, \bibinfo{person}{Cong~Chi
  Tran}, \bibinfo{person}{Yu Chen}, \bibinfo{person}{Ke Tan}, {and}
  \bibinfo{person}{Jean~Luc Habiyaremye}.} \bibinfo{year}{2017}\natexlab{}.
\newblock \showarticletitle{Effect of user interface layout on the operators’
  mental workload in emergency operating procedures in nuclear power plants}.
\newblock \bibinfo{journal}{\emph{Nuclear Engineering and Design}}
  \bibinfo{volume}{322} (\bibinfo{year}{2017}), \bibinfo{pages}{266--276}.
\newblock
\showISSN{0029-5493}
\urldef\tempurl%
\url{https://doi.org/10.1016/j.nucengdes.2017.07.012}
\showDOI{\tempurl}


\bibitem[Yazgan et~al\mbox{.}(2021)]%
        {yazgan2021overview}
\bibfield{author}{\bibinfo{person}{Ebru Yazgan}, \bibinfo{person}{SERT Erdi},
  {and} \bibinfo{person}{Deniz {\c{S}}{\.I}M{\c{S}}EK}.}
  \bibinfo{year}{2021}\natexlab{}.
\newblock \showarticletitle{Overview of Studies on the Cognitive Workload of
  the Air Traffic Controller}.
\newblock \bibinfo{journal}{\emph{International Journal of Aviation Science and
  Technology}} \bibinfo{volume}{2}, \bibinfo{number}{01}
  (\bibinfo{year}{2021}), \bibinfo{pages}{28--36}.
\newblock


\bibitem[Zhang et~al\mbox{.}(2020)]%
        {zhang2020study}
\bibfield{author}{\bibinfo{person}{Jingling Zhang}, \bibinfo{person}{Daizhong
  Su}, \bibinfo{person}{Yan Zhuang}, {and} \bibinfo{person}{QIU Furong}.}
  \bibinfo{year}{2020}\natexlab{}.
\newblock \showarticletitle{Study on cognitive load of OM interface and eye
  movement experiment for nuclear power system}.
\newblock \bibinfo{journal}{\emph{Nuclear Engineering and Technology}}
  \bibinfo{volume}{52}, \bibinfo{number}{1} (\bibinfo{year}{2020}),
  \bibinfo{pages}{78--86}.
\newblock


\bibitem[Zhang et~al\mbox{.}(2017a)]%
        {zhang2017s}
\bibfield{author}{\bibinfo{person}{Xucong Zhang}, \bibinfo{person}{Yusuke
  Sugano}, \bibinfo{person}{Mario Fritz}, {and} \bibinfo{person}{Andreas
  Bulling}.} \bibinfo{year}{2017}\natexlab{a}.
\newblock \showarticletitle{It's written all over your face: Full-face
  appearance-based gaze estimation}. In \bibinfo{booktitle}{\emph{Computer
  Vision and Pattern Recognition Workshops (CVPRW), 2017 IEEE Conference on}}.
  IEEE, \bibinfo{pages}{2299--2308}.
\newblock


\bibitem[Zhang et~al\mbox{.}(2017b)]%
        {zhang2017mpiigaze}
\bibfield{author}{\bibinfo{person}{Xucong Zhang}, \bibinfo{person}{Yusuke
  Sugano}, \bibinfo{person}{Mario Fritz}, {and} \bibinfo{person}{Andreas
  Bulling}.} \bibinfo{year}{2017}\natexlab{b}.
\newblock \showarticletitle{Mpiigaze: Real-world dataset and deep
  appearance-based gaze estimation}.
\newblock \bibinfo{journal}{\emph{IEEE transactions on pattern analysis and
  machine intelligence}} \bibinfo{volume}{41}, \bibinfo{number}{1}
  (\bibinfo{year}{2017}), \bibinfo{pages}{162--175}.
\newblock


\bibitem[Zhang et~al\mbox{.}(2004)]%
        {zhang2004driver}
\bibfield{author}{\bibinfo{person}{Yilu Zhang}, \bibinfo{person}{Yuri Owechko},
  {and} \bibinfo{person}{Jing Zhang}.} \bibinfo{year}{2004}\natexlab{}.
\newblock \showarticletitle{Driver cognitive workload estimation: A data-driven
  perspective}. In \bibinfo{booktitle}{\emph{Proceedings. The 7th International
  IEEE Conference on Intelligent Transportation Systems (IEEE Cat. No.
  04TH8749)}}. IEEE, \bibinfo{pages}{642--647}.
\newblock


\bibitem[Ziv(2016)]%
        {ziv2016gaze}
\bibfield{author}{\bibinfo{person}{Gal Ziv}.} \bibinfo{year}{2016}\natexlab{}.
\newblock \showarticletitle{Gaze behavior and visual attention: A review of eye
  tracking studies in aviation}.
\newblock \bibinfo{journal}{\emph{The International Journal of Aviation
  Psychology}} \bibinfo{volume}{26}, \bibinfo{number}{3-4}
  (\bibinfo{year}{2016}), \bibinfo{pages}{75--104}.
\newblock


\end{thebibliography}


\end{document}